\documentclass[a4paper,11pt]{article}
\pdfoutput=1
\usepackage{jcappub}
\usepackage{bm}
\usepackage{soul}
\usepackage{latexsym}
\usepackage{dcolumn}
\usepackage{amsfonts,amssymb,amsmath}
\usepackage{graphicx,epsfig}
\usepackage{psfrag}
\usepackage{subfigure}
\usepackage{rotating}
\usepackage{hyperref}
\usepackage{tikz}
\hypersetup{
	unicode=false,          % non-Latin characters in Acrobat’s bookmarks
	pdftoolbar=true,        % show Acrobat’s toolbar?
	pdfmenubar=true,        % show Acrobat’s menu?
	pdffitwindow=false,     % window fit to page when opened
	pdfstartview={FitH},    % fits the width of the page to the window
	pdftitle={My title},    % title
	pdfauthor={Author},     % author
	pdfsubject={Subject},   % subject of the document
	pdfcreator={Creator},   % creator of the document
	pdfproducer={Producer}, % producer of the document
	pdfkeywords={keyword1} {key2} {key3}, % list of keywords
	pdfnewwindow=true,      % links in new window
	colorlinks=true,       % false: boxed links; true: colored links
	linkcolor=red,          % color of internal links
	citecolor=cyan,        % color of links to bibliography
	filecolor=magenta,      % color of file links
	urlcolor=blue,           % color of external links
	linktocpage=true
}

\newcommand{\beq}{\begin{equation}}
\newcommand{\eeq}{\end{equation}}
\newcommand{\bal}{\begin{aligned}}
	\newcommand{\eal}{\end{aligned}}
\newcommand{\bea}{\begin{eqnarray}}
\newcommand{\eea}{\end{eqnarray}}

\newcommand{\dd}{\mathrm{d}}

\newcommand{\fatdot}{\raisebox{0.25ex}{\tikz\filldraw[black,x=2pt,y=2pt] (0,0) circle (0.2);}}

\def\beq{\begin{equation}}
\def\efq{\end{equation}}
\def\br{\begin{eqnarray}}
\def\er{\end{eqnarray}}
\def\benu{\begin{enumerate}}
	\def\efnu{\end{enumerate}}
\def\nn{\nonumber}

\def\l{\left}
\def\r{\right}
\def\cR{{\cal R}}
\def\d{{\rm d}}
\def\f{\frac}
\def\nn{\nonumber} 

\def\cR{{\mathcal{R}}}
\def\cS{{\mathcal{S}}}

\def\Mpl{{M_{_{\textup{Pl}}}}}

\def\Mpl{M_{_{\rm Pl}}}

\def\d{{\rm d}}

\makeatletter
\gdef\@fpheader{}
\makeatother

\begin{document}
\title{Probing Primordial Features with the Stochastic Gravitational Wave Background}

\author[1,2,3]{Matteo~Braglia,}
\author[4]{Xingang Chen,}
\author[5,2]{Dhiraj Kumar Hazra}

	\affiliation[1]{DIFA, Dipartimento di Fisica e Astronomia,
		Alma Mater Studiorum, Universit\`a degli Studi di Bologna,
		Via Gobetti, 93/2, I-40129 Bologna, Italy}
   \affiliation[2]{INAF/OAS Bologna, Osservatorio di Astrofisica e Scienza dello Spazio,
		Area della ricerca CNR-INAF, via Gobetti 101, I-40129 Bologna, Italy}
	\affiliation[3]{INFN, Sezione di Bologna,
		via Irnerio 46, I-40126 Bologna, Italy}
	
\affiliation[4]{Institute for Theory and Computation, Harvard-Smithsonian Center for Astrophysics, 60
Garden Street, Cambridge, MA 02138, USA}
\affiliation[5]{The  Institute  of  Mathematical  Sciences,  HBNI,  CIT  Campus, Chennai  600113,  India}

\emailAdd{matteo.braglia2@unibo.it, xingang.chen@cfa.harvard.edu, dhiraj@imsc.res.in}

	\abstract
{The stochastic gravitational wave background (SGWB) offers a new opportunity to observe signals of primordial features from inflationary models.
We study their detectability with future space-based gravitational waves experiments, focusing our analysis on the frequency range of the LISA mission.
We compute gravitational wave spectra from primordial features by exploring the parameter space of a two-field inflation model capable of generating different classes of features.
Fine-tuning in scales and amplitudes is necessary for these signals to fall in the observational windows. In some cases the scalar power spectrum can significantly exceed the $n_s=5$ limit in single-field inflation and grow as fast as $n_s=9.1$. 
Once they show up, several classes of frequency-dependent oscillatory signals, characteristic of different underlying inflationary physics, may be distinguished and the SGWB provides a window on dynamics of the primordial universe independent of cosmic microwave background and large-scale structure. To connect with future experimental data, we discuss two approaches of how the results may be applied to data analyses. 
First, we discuss the possibility of reconstructing the signal with LISA, which requires a high signal-to-noise ratio. The second more sensitive approach is to apply templates representing the spectra as estimators. 
For the latter purpose, we construct templates that can accurately capture the spectral features of several classes of feature signals and compare them with the SGWB produced by other physical mechanisms.
}

	\maketitle
	
\section{Introduction}
\label{sec:introduction}

As the leading candidate theory for the origin of the Big Bang, the inflation scenario may have operated at an energy scale far beyond the reach of any terrestrial experiments. At higher energies, we expect the existence of many new particles with heavier masses and richer interactions than those known in the Standard Model. During the inflationary epoch, the inflaton should be rolling along a path embedded within a sophisticated potential landscape formed by these new particles and interactions. One may envision many natural ways such a rolling can proceed beyond the simplest scenario of a smooth single field inflation. One such possibility is that the inflaton deviates from the smooth rolling temporarily from time to time. Physics that cause these disturbances are generally referred to as ``primordial features".

According to the temporal characteristics of the primordial features, there are roughly two categories in which these features can impact the dynamics of the inflaton and the predictions of the primordial density perturbations -- features that proceed in time intervals larger than the Hubble time or in complete isolation, and features that proceed as a series of perturbations with temporal frequencies higher than the Hubble rate.

In the first category, at the leading order, the effects of the features may be approximated as a sum of those of individual ones. Each individual feature is called a ``sharp feature".
A single sharp feature induces a temporally localized perturbation on the rolling of the inflaton and affects its quantum fluctuations that have wavelengths similar to, or shorter than, the horizon size at that moment. These perturbed quantum modes lead to a sinusoidally wavenumber-dependent deviation from the scale-invariant density perturbations \cite{Starobinsky:1992ts}. This scale-dependence is commonly referred to as the ``sinusoidal running". To clarify the two closely related nomenclatures, note that, while the sharp feature refers to the primordial feature present in the physical model, the sinusoidal feature is often used to refer to the feature in the primordial power spectrum (PPS) generated by the sharp feature in the model.

In the second category, the temporal frequency of the features is higher than the Hubble rate and their collective contribution to the density perturbations becomes important. These high frequency background oscillations resonate with the quantum fluctuations of the inflaton (or, more generally, any field sourcing the density perturbations) while the fluctuations are still at the sub-horizon scales, mode by mode \cite{Chen:2008wn}. This type of features is called the ``resonant features". If the time-dependence of the resonant features is exactly sinusoidal, as often used as the prototype example, these perturbations lead to a sinusoidal dependence on the {\em logarithm} of wavenumbers, commonly referred to as the ``resonant running".

Each category can have many different realizations in terms of physical models. For example, sharp feature signals may be produced by any isolated feature, such as a kink, step or bump, in the inflationary potential \cite{Starobinsky:1992ts,Adams:2001vc,Chen:2006xjb,Hazra:2010ve,Adshead:2011jq,Hazra:2014goa} or the internal field space such as the sound speed \cite{Bean:2008na,Miranda:2012rm}. Sharp feature signals can also be induced by a more complicated change in a multifield model, in which, e.g., the inflaton takes a sharp turn \cite{Achucarro:2010da,Chen:2011zf,Gao:2012uq} or undergoes a tachyonic falling \cite{Chen:2014joa,Chen:2014cwa}. Resonant feature signals may be generated by periodic ripples in the inflaton potential or internal space \cite{Chen:2008wn,Flauger:2009ab,Flauger:2010ja,Chen:2010bka,Aich:2011qv}. They may also be generated by classically oscillating massive fields that are perturbed out of their equilibrium points by sharp features \cite{Chen:2011zf,Chen:2011tu,Chen:2014joa,Chen:2014cwa}. These realizations create a variety of feature profiles in the density perturbations. Even within the same category where they share the similar sinusoidal or resonant running, the oscillatory features can have different scale-dependent {\em envelops} in different models.

There are also primordial feature models that combine the characteristics of both categories. For example, in one of the examples mentioned above, when a massive field gets temporarily perturbed out of its equilibrium point, the sharp feature that creates this perturbation and the subsequent oscillations of the massive field generate two sets of signals, with the characters of sharp feature and resonant feature, respectively, which are smoothly connected with each other \cite{Chen:2011zf,Chen:2011tu,Chen:2014joa,Chen:2014cwa}. This is referred to as the ``standard clock signal".

Signals of primordial features, if detected, could be used to probe a wide range of physics in the primordial universe, including shapes of inflationary potentials, properties of new particles and direct evidences for inflation. See \cite{Chen:2010xka,Chluba:2015bqa,Slosar:2019gvt} for reviews.
Prospects of observing these signals have been extensively studied using cosmic microwave background (CMB) \cite{Akrami:2018odb,Hazra:2017joc}, large-scale structure (LSS) \cite{Huang:2012mr,Hazra:2012vs,Chen:2016vvw,Ballardini:2016hpi,Palma:2017wxu,LHuillier:2017lgm,Ballardini:2017qwq,Beutler:2019ojk,Ballardini:2019tuc,Debono:2020emh} and 21-cm hydrogen line experiments \cite{Chen:2016zuu,Xu:2016kwz}. 

In this paper, we explore another opportunity of observing primordial features through the stochastic gravitational wave background (SGWB) experiments. Primordial scalar perturbations generated during inflation act as a source of gravitational waves when they re-enter the horizon during the radiation-dominated epoch \cite{Matarrese:1993zf,Matarrese:1997ay,Acquaviva:2002ud,Carbone:2004iv,Ananda:2006af,Baumann:2007zm,Assadullahi:2009nf}. If the imprints of primordial features in the scalar perturbations have large enough amplitudes, the gravitational waves they source may be detected by the SGWB experiments. However, current SGWB experiments are sensitive to features around $\sim 20-25$  $e$-folds before the end of inflation\footnote{The exact number of $e$-folds depends on the specific mechanism producing the SGWB.}, which are located at much larger wavenumbers compared to those studied in the CMB, LSS and 21-cm experiments. At the same time, their oscillatory amplitudes have to be significantly larger to have any chance of being detected.
From the model-building point of view, these two requirements are nonetheless consistent with each other: breaks from slow-roll are allowed to be more dramatic if fewer subsequent e-folds of inflation is needed. The existence of such features, however, is not found to be natural at all, and fine-tuning in scale-location and signal-amplitude is needed for them to become observable and satisfy observational bounds.
From the observational point of view, since signals of many feature models are transient and disappear in a few $e$-folds, primordial features that could possibly be probed by SGWB experiments are independent from those by the other experiments mentioned above.

We start out in Sec.~\ref{sec:Model} and App.~\ref{app:numericalmethod} by reviewing a two-field inflation model that contains both a sharp feature and resonant features. This model is originally constructed for the purpose of studying classical primordial standard clocks \cite{Chen:2014joa,Chen:2014cwa}. However, as we vary the model parameters, the relative significance between the two categories of signals change smoothly, and we are led to many different types of features reviewed above. These include sharp features signals with different envelop behaviors, resonant features, and features of primordial standard clocks which are combinations of both features. So, we will use this model as the prototype models for a wide range of primordial features, and study their background evolution and scalar power spectra in Sec.~\ref{sec:ModelI}, \ref{sec:ModelII} and \ref{sec:ModelIII}. The SGWB signals generated by these features are computed numerically in Sec.~\ref{sec:SGWBandDetect}. To study the experimental prospects of these features, we discuss two approaches: in Sec.~\ref{sec:reconstruction}, the approach of signal reconstruction, in which we use the so-called binned power-law sensitivity-curve technique to discuss qualitatively to what extent different classes of signals may be detected and characterized by future gravitational wave interferometers; and in Sec.~\ref{sec:temp_est}, the approach of template estimators, in which we provide simple analytical templates that summarize the main properties of various gravitational wave profiles obtained in this paper, which can be easily used by the gravitational wave (GW) community.

\section{A two-field model and useful formulae}
\label{sec:Model}

	In this section, we briefly review a two-field inflation model that contains both sharp and resonant feature \cite{Chen:2014joa,Chen:2014cwa}. The relative significance between the two can be tuned by exploring different parts of the parameter space, leading to a wide variety of phenomena.
	
	This model has two stages of slow-roll inflation connected by features that temporarily perturb the slow-roll inflationary background. The sharp feature is represented by a tachyonic falling of the inflaton into a potential well, perturbing the inflaton away from the slow-roll. After that, the inflaton oscillates and settles down in the massive field direction of the potential well, and at the same time, finds the tangential slow-roll direction along the valley of this potential well and starts the second stage of slow-roll inflation. This may be one of possible ways the inflaton searches for the rare slow-roll directions in a complicated landscape. The tachyonic potential generates the sharp feature signal and the oscillation of the massive field generates the resonant feature signal. Which signals are more important depends on the shape of the potential and the mass and couplings of the massive field.\footnote{Note that, in the parameter space that is of interest in Ref.~\cite{Chen:2014joa,Chen:2014cwa}, the background is still inflating during the time of the sharp and resonant features, although it deviates a little from the slow-roll inflation. The tachyonic falling and massive field oscillation are only small perturbations to the leading properties of inflation. This is because CMB and LSS constraints in the large scales only allow perturbative features. In this paper, in order to make feature signals accessible to GW experiments, in some cases, the effects of the features are so large such that the background expansion may temporarily deviate from inflation, entering a short period of matter-domination epoch. These cases are similar to models considered in \cite{Polarski:1992dq,Polarski:1994rz,Polarski:1995zn}.}

The Lagrangian of the feature model is given by \cite{Chen:2014cwa,Chen:2014joa}:
	\begin{equation}
\mathcal{L}=-\frac{1}{2}f^2\left(\sigma\right) g^{\mu\nu}\partial_\mu\theta\partial_\nu\theta-U(\Theta)-\frac{g^{\mu\nu}}{2}\partial_\mu\sigma\partial_\nu\sigma-V(\sigma),
\label{eq:LagrOrig}
	\end{equation}
where we assume the form $f(\sigma)=1+\xi\sigma$ for the non-canonical coupling, unless stated otherwise.

The relevant equations to study the background evolution are the Klein-Gordon equations for the two scalar fields
\begin{subequations}
		\label{eq:sf}
		\begin{eqnarray}
		\label{eq:KGphi}
		\ddot{\sigma}+3 H\dot{\sigma}+V_\sigma
		&=&f_\sigma f \dot{\Theta}^2,\\
		\label{eq:KGchi}
		\ddot{\Theta}+\left(3 H+ 2 \frac{f_\sigma}{f} \dot{\sigma}\right) \dot{\Theta}
		+\frac{U_\Theta}{f^2}&=&0 ~,
		\end{eqnarray}
	\end{subequations}
and the Friedmann equations
	\begin{subequations}
		\label{eq:f}
		\begin{eqnarray}
		H^2&=&\frac{1}{3 \Mpl^2}		\left[\frac{\dot{\sigma}^2}{2}+	V	+f^2\frac{\dot{\Theta}^2}{2}+U\right],\\
		\dot{H}&=&-\frac{1}{2 \Mpl^2}\left[\dot{\sigma}^2
		+f^2\dot{\Theta}^2\right].
		\end{eqnarray}
	\end{subequations}
	
The potential $U(\Theta)$ is a slow-roll potential that always satisfies the slow-roll conditions. The potential $V(\sigma)$ contains a potential well which is the sharp feature that triggers the tachyonic falling.
We are interested in the case in which inflation is first supported by the scalar field $\sigma$, which eventually falls into the bottom of its potential dip, and  oscillates for a while violating the slow-roll conditions, before settling down to the minimum of the potential at $\sigma=0$. Then the second stage of inflation, which lasts around $\sim20$ $e$-folds in the specific case considered in our paper, consists in an effective single field inflation driven by $\Theta$. 
Note that the  oscillation frequency is determined by the effective mass of the  \emph{clock} field  $V_{\sigma\sigma}(\sigma=0)$, and the evolution of the oscillation amplitude is typical of the massive field $a^{-3/2}$ simply due to the spatial dilution of the inflationary background.

In order to source a SGWB which can be detected by GW interferometers, a large amplification of curvature (and hence density) perturbations is needed. In this paper, for simplicity, we focus on space-based GW experiments of which LISA, BBO and DECIGO are representative examples. Typically, the amplitude of the power spectrum should be ${\cal O}(10^{-3})$ in order for them to fall within the observational threshold of LISA, or ${\cal O}(10^{-4})$ to fall within the sensitivity ranges of BBO and DECIGO. This means that the tachyonic feature introduced in $V(\sigma)$ has to be much more significant than those studied in the context of CMB; but as we shall see, this is not difficult to achieve.

\begin{figure}
		\begin{center} 
			\resizebox{214pt}{172pt}{\includegraphics{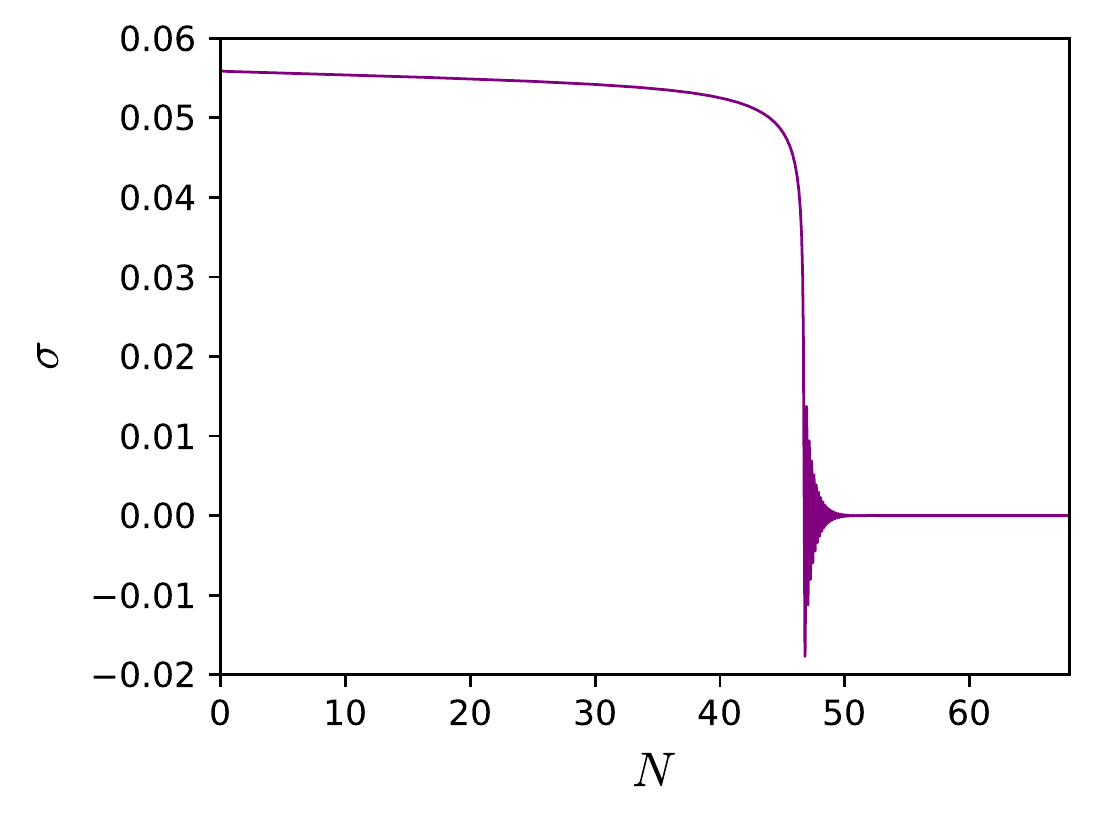}}
			\resizebox{214pt}{172pt}{\includegraphics{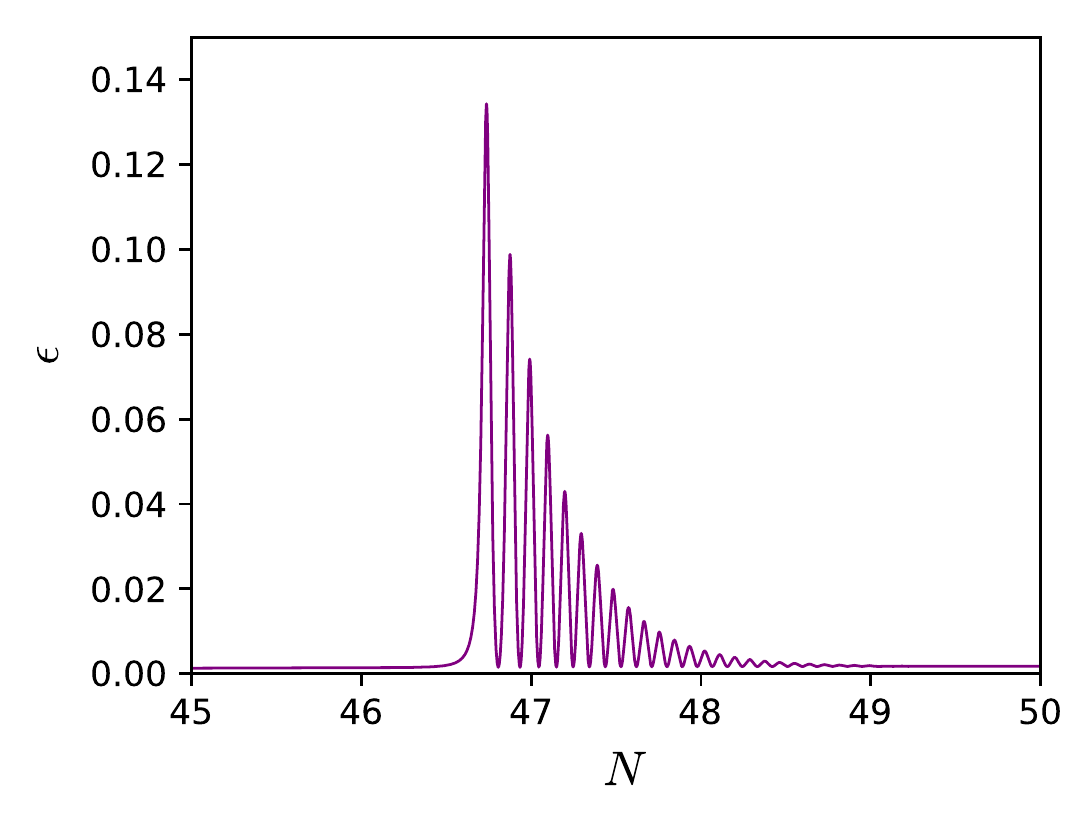}}	
		\end{center}
		\caption{\label{fig:backgroundIntro} 
		Evolution of the clock field  [left] and the first slow-roll parameter $\epsilon$ [right] for Model I in Eq.~\eqref{eq:potI}. The parameters used in this plot are $C_\sigma=0.02,\,$ $\sigma_f=0.0164,\,$ $C_\Theta=0.01,\,$  $\sigma_i=0.0556,\,$ and $\Theta_i=3.16$. }
	\end{figure}

	An example of the background evolution for our model is shown in Fig.~\ref{fig:backgroundIntro}, where we plot the evolution of the clock field $\sigma$  and the first slow-roll parameter $\epsilon$ using the parameters given in the caption. For definiteness, we used the Model I presented in Section~\ref{sec:ModelI} and described by the potential in Eq.~\eqref{eq:potI}.  
	
  As discussed in Ref.~\cite{Braglia:2020eai} (see also Ref.~\cite{Renaux-Petel:2015mga}), if the non-canonical coupling is large enough, the mass squared of isocurvature modes becomes negative and isocurvature perturbations experience a transient tachyonic growth, amplifying the curvature perturbations around the scales that cross the horizon during slow-roll violation. In this paper, we make use of the full numerical pipeline, developed in Ref.~\cite{Braglia:2020eai}, to compute the PPS, the SGWB density and the abundance of PBHs. 
	
	The specific details of the second stage of inflation and consequences on the shape of the PPS and SGWB will be the subject of the following sections.
In the following, we always solve for $\sigma_i$ to get the maximum violation of the slow-roll around $\sim20$ $e$-folds before the end of inflation, so that the peak of the power spectrum occurs at scales where LISA has its maximum sensitivity~\cite{Bartolo:2018evs}. Also note that we always make sure that the maximum peak in the power spectrum is $\mathcal{P}_{\mathcal{R}\,,{\rm max}}<0.01$ in order to avoid production of PBH that can contribute significantly to the cold dark matter density and is tightly constrained by observations \cite{Carr:2020gox}. 

In the next sections, to compute the PPS, we follow the perturbation theory and numerical integration outlined in Appendix~\ref{app:numericalmethod}  and make use of a two-field extension of the code BINGO~\cite{Hazra:2012yn}. 
The density perturbations are computed numerically by solving the coupled differential equations of two perturbed fields \cite{Tsujikawa:2002qx,Lalak:2007vi}. In terms of the in-in formalism, this procedure is equivalent to summing up all tree-level diagrams for the two-point function non-perturbatively \cite{Chen:2015dga}. Significance of loop corrections would have to be determined by the strength of model-dependent non-linear couplings in the inflation model, which we leave for future study.
For simplicity, we will work in units where $M_{\rm pl}=1$.

%%%%%%%%%%%%%%%%%%%%%%%%%%%%%%%%%%%%%%%%%%%%%%%%%%%%%%%%%%%%%%%%%%%%%%%%%%%%%%%
	
\section{Model variation and feature profiles}
\label{sec:ModelIntro}

In this section, we vary the model parameters and numerically compute primordial feature profiles in the density perturbations. 
In different parts of the parameter space, the relative importance of the two categories of features present in this model can change smoothly; the envelopes of the oscillatory signals in the momentum space also vary model-dependently. This results in a rich spectrum of predictions, representing broad classes of primordial feature models that we would like to study. For later convenience, we will divide these features into the following four classes, named by their most important characters:

\begin{itemize}
\item {\em Bump feature}. This in principle belongs to the sharp feature signal generated by a sharp feature. However, the first bump in its sinusoidal running is significantly larger than the rest of the bumps which effectively become negligible.

\item {\em Sinusoidal feature}. This refers to the typical sharp feature signal generated by a sharp feature, as reviewed in Sec.~\ref{sec:introduction}. The feature exhibits a sinusoidal running with a model-dependent envelope.

\item {\em Resonant feature}. This refers to the typical resonant feature signal generated by high frequency periodic features, as reviewed in Sec.~\ref{sec:introduction}. The feature exhibits a resonant running with a model-dependent envelope.

\item {\em Standard clock feature}. This refers to the feature generated by a classically excited, oscillating massive field. This feature is a smooth combination of sinusoidal feature (in the larger scales) and resonant feature (in the shorter scales).
As we shall see later, typically it is difficult for both of them to appear in the SGWB spectrum, and only the more significant one shows up.

\end{itemize}

\subsection{Model I}
\begin{figure}
		\begin{center} 
			\resizebox{214pt}{172pt}{\includegraphics{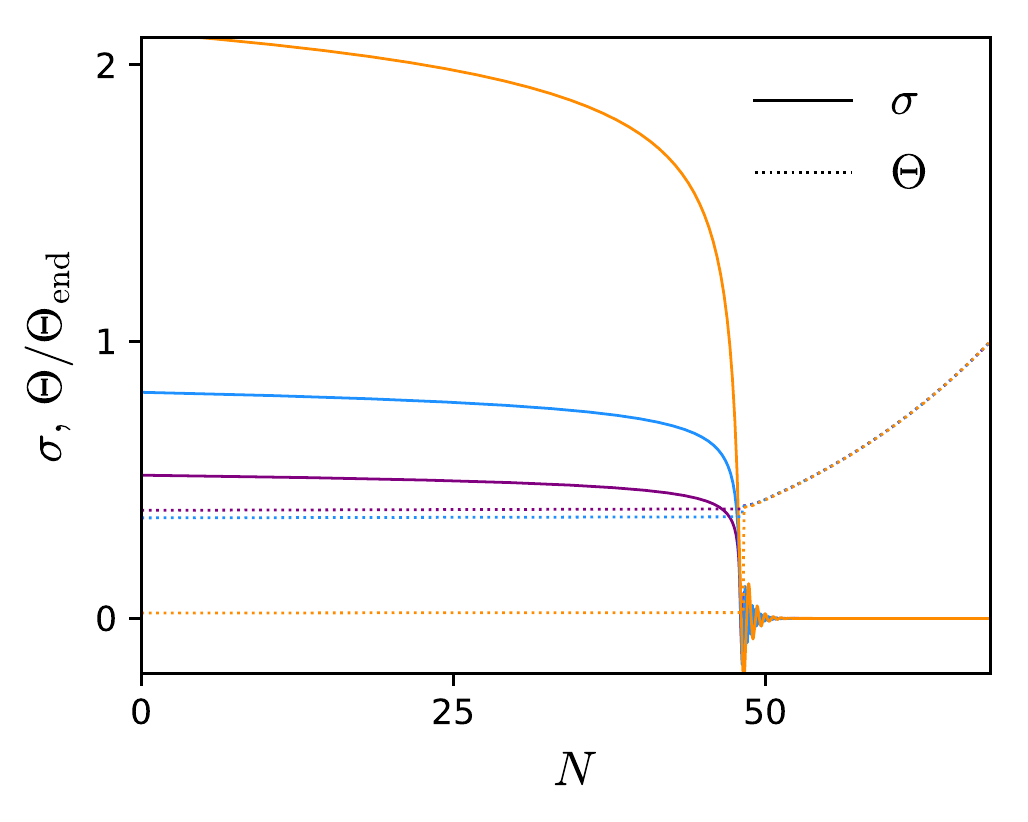}}
			\resizebox{214pt}{172pt}{\includegraphics{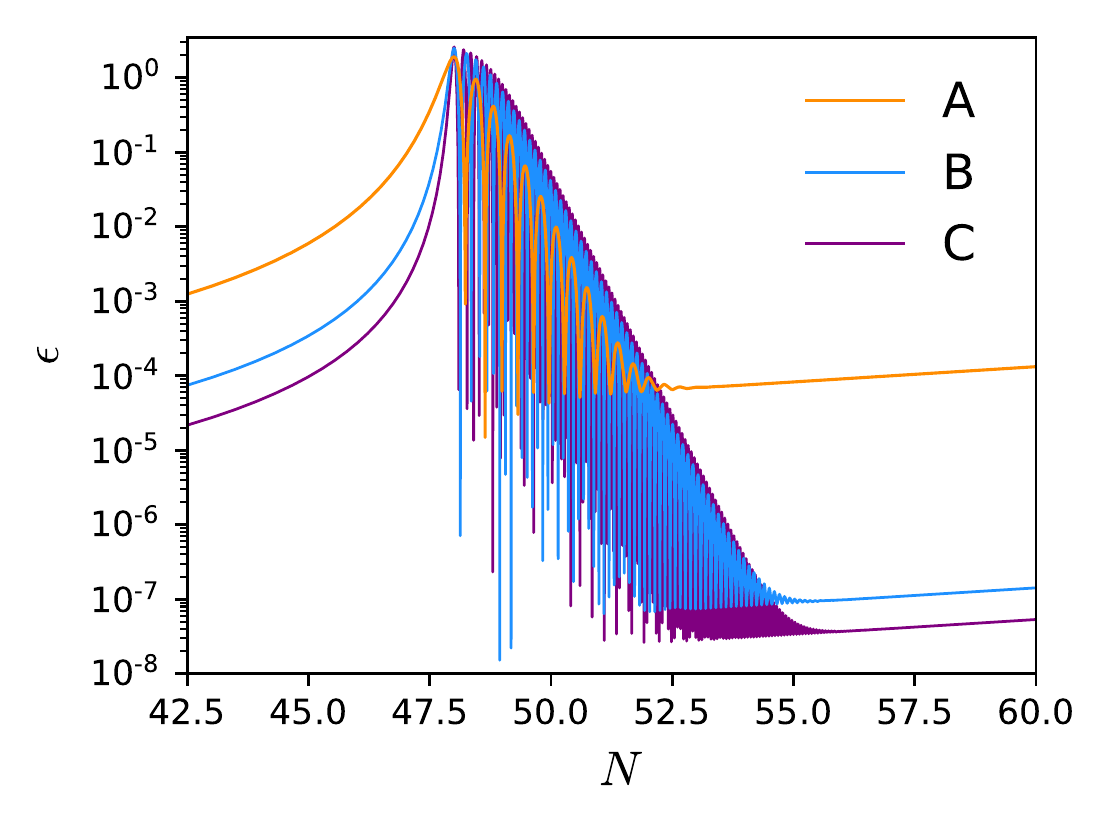}}	
		\end{center}
		\caption{\label{fig:BGModelI} Evolution of the two scalar fields [left] and of the $\epsilon$ parameter [right] for Model I. The parameters used in the three examples are A:$\{V_0=3.99\times10^{-13}\,, \xi\sigma_f=3.32,\,\sigma_f=0.775,\,\sigma_i=2.12,\,\Theta_i=9.84\times10^{-3}\}$, B:$\{V_0=2.84\times10^{-14}\,, \xi\sigma_f=1.45,\,\sigma_f=0.258,\,\sigma_i=0.817,\,\Theta_i=6.01\times10^{-3}\}$ and C:$\{V_0=8.62\times10^{-15}\,, \xi\sigma_f=1.086,\,\sigma_f=0.155,\,\sigma_i=0.517,\,\Theta_i=3.96\times10^{-3}\}$. For all three cases we fix $C_\sigma=10$ and $C_\Theta=0.0476$. }
	\end{figure}

%\section{Model I: small field inflation in the $\Theta$ direction}
\label{sec:ModelI}
As a first example, we consider the following model, introduced in Refs.~\cite{Chen:2014cwa,Chen:2014joa}:
\begin{equation}
\label{eq:potI}
U(\Theta)=V_0\left(1- C_\Theta\Theta^2\right),\quad
	V(\sigma)=V_0\,C_\sigma\,\left[1-\exp\left(-\sigma^2/\sigma_f^2\right)\right],
\end{equation}
where we have normalized both the potentials to the cosmological constant term $V_0$ for simplicity. We plot the background evolution for three selected examples in Fig.~\ref{fig:BGModelI}. Note that, as discussed in the previous Section, the frequency of the massive field oscillations depends on $\sigma_f$ and increases with its effective mass  $m_\sigma/H=\sqrt{2 C_\sigma}/\sigma_f$, where $H$ is the Hubble parameter during the second stage of inflation. Note that, with the potential $U(\Theta)$ in Eq.~\eqref{eq:potI}, the second inflationary stage in the   $\Theta$ direction is a small field one. The $\epsilon$ parameter during this  stage can therefore take very small values. The end of the inflation is achieved by cutting off the potential $U(\Theta)$ with a hybrid field, and is not a concern of this paper because there are many ways to realize it.

As discussed in Ref.~\cite{Palma:2020ejf,Fumagalli:2020adf,Braglia:2020eai}, a large  non-canonical coupling between the two scalar field is needed to generate large curvature perturbations. Because of that, the potential driving term $U_\Theta$ in the Klein-Gordon equation for $\Theta$ is suppressed by a factor of $f(\sigma)^{-2}$ and $\Theta$ stays frozen to its initial condition until the clock field starts to oscillate. 

\begin{figure}
		\begin{center} 
			\includegraphics[width=\columnwidth]{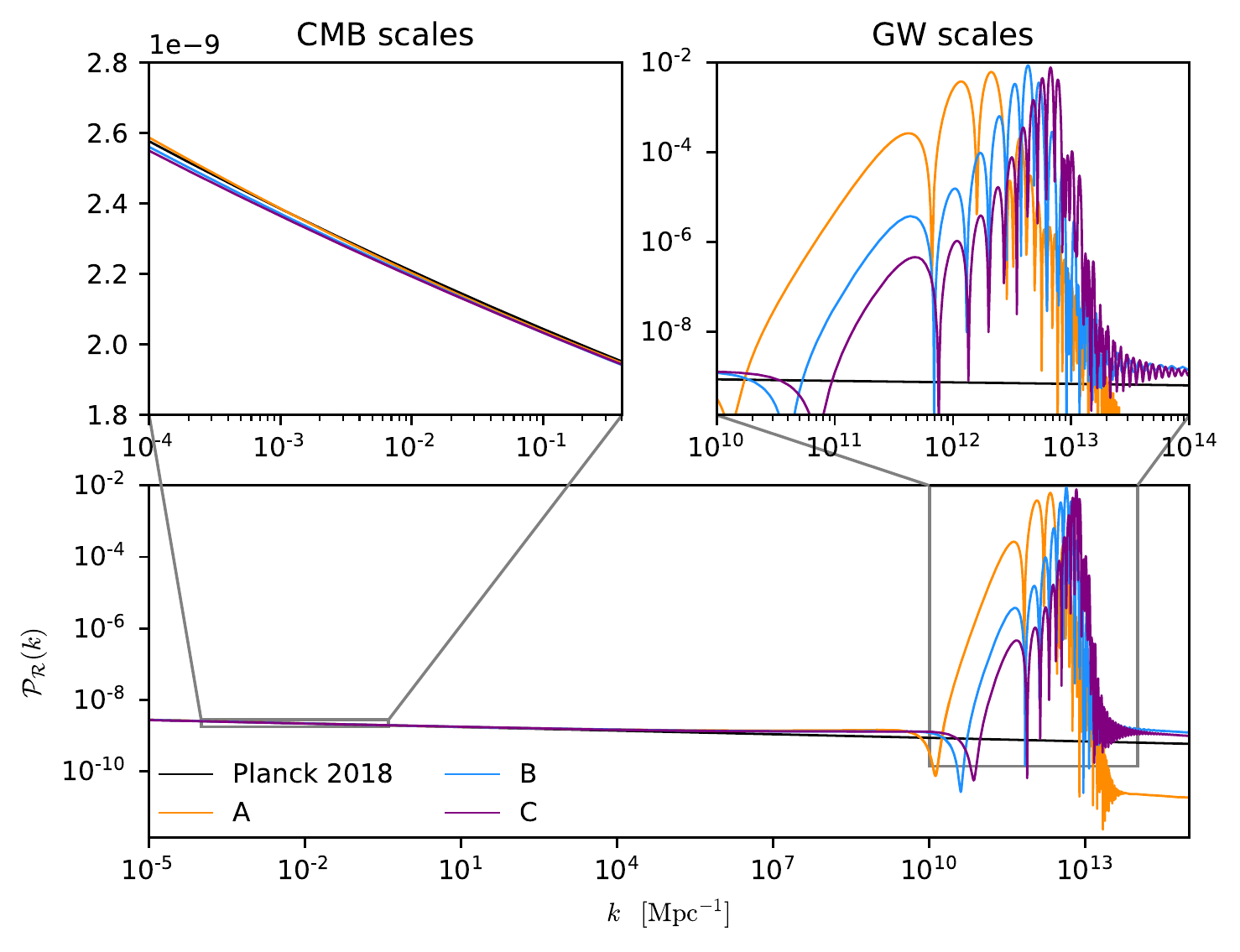}	
		\end{center}
		\caption{\label{fig:PkModelI} PPS [bottom] zoomed at CMB [top-left] and LISA [top-right] scales  for Model I. The parameters used in the three examples are the one reported in the caption of Fig.~\ref{fig:BGModelI}. For a comparison, we also plot the Planck 2018 power-law spectrum with $n_s=0.9665$  and $A_s=2.09\times 10^{-9}$ at the pivot scale $k=0.05\, {\rm Mpc}^{-1}$ \cite{Akrami:2018odb}. }
	\end{figure}

We present the PPS obtained for these examples in Fig.~\ref{fig:PkModelI}, zooming in the CMB and LISA scales. Note that the spectral indices computed at the pivot scale $k=0.05\,{\rm Mpc}^{-1}$, $n_s=0.9654,\, 0.9666,\,0.9672$ for the cases denoted by A, B and C respectively, are completely consistent with the Planck constraints of $n_s=0.9665\pm0.0038$ at 68 \% CL for Planck TT,TE,EE + lowE + lensing + BAO \cite{Akrami:2018odb}. In fact, given the strong coupling regime, i.e. large $\xi\sigma_f$, isocurvature sourcing can also be important at CMB scales and can modify $n_s$. The latter, can always be adjusted by tuning the parameters of the model. We also emphasize that choosing a different form for the potential of the $V(\sigma)$ can  also change $n_s$.

At smaller scales, the standard clock feature generated by the oscillations of the scalar field is amplified by the large non-canonical coupling and reaches an amplitude as large as $\mathcal{P}_\mathcal{R}(k_{\rm peak})\lesssim\mathcal{O}(0.01)$. Going from larger to smaller scales, the signal is composed by the sharp feature with a sinusoidal running which consists in the broader bumps before the point of maximum amplitude, after which we see the character of the resonant feature. The resonant feature contains some unusual modulation envelope, because, as can be seen from Fig.~\ref{fig:BGModelI}, the intermediate stage during which the massive field oscillates is very briefly matter dominated, i.e.~$\epsilon\gtrsim1$, causing the new phenomenology observed here. Eventually, the amplitude of the peaks decays and at very smaller scales we observe a plateau  produced during the second stage of inflation by the field $\Theta$ and a superimposed oscillatory pattern with the characteristic running of the pure resonant feature.

The effect of increasing the frequency of the clock field is to intensify the number of peaks and to \emph{squeeze} the power spectrum around the peak of maximum amplitude, decreasing the amplitude of the ones at larger scales, particularly the first bump.

%%%%%%%%%%%%%%%%%%%%%%%%%%%%%%%%%%%%%%%%%%%%%%%%%%%%%%%%%%%%%%%%%%%%%%%%%%%%%%%

\subsection{Model II}
%\section{Model II: $m^2\,\sigma^2$ correction to Model I}
\label{sec:ModelII}	
As a second example, we consider the following modification by changing the shape of the potential in the $\sigma$ direction, also introduced in Refs.~\cite{Chen:2014cwa}:
\begin{equation}
\label{eq:potII}
U(\Theta)=V_0\left(1- C_\Theta\Theta^2\right),\quad
	V(\sigma)=V_0\,C_\sigma\,\left[1-\exp\left(-\sigma^2/\sigma_f^2\right)\right]+\frac{\left(m_0\sqrt{V_0}\right)^2}{2}\sigma^2.
\end{equation}

\begin{figure}
		\begin{center} 
			\resizebox{214pt}{172pt}{\includegraphics{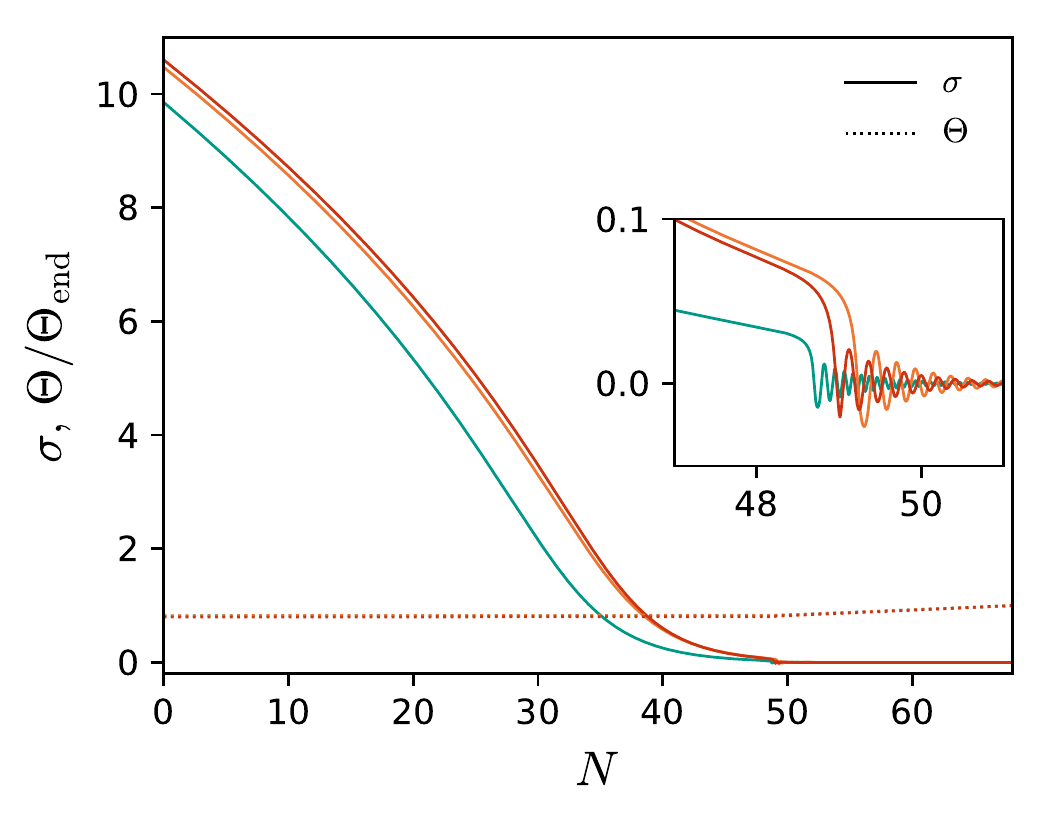}}
			\resizebox{214pt}{172pt}{\includegraphics{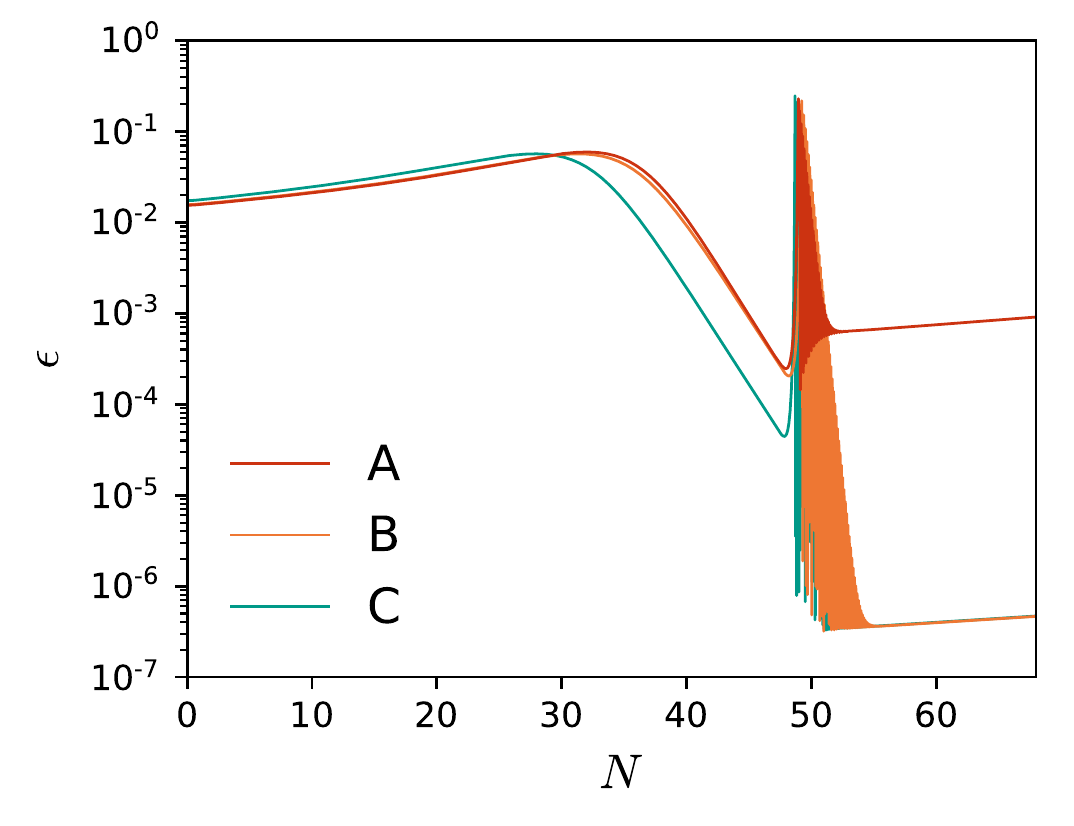}}	
		\end{center}
		\caption{\label{fig:BGModelII} Evolution of the two scalar fields [left] and of the $\epsilon$ parameter [right] for Model II. The parameters used in the three examples are 
		A: $\{V_0=1.86\times10^{-9}\,, \xi\sigma_f=1.15,\,\sigma_f=0.0258,\,\sigma_i=10.6,\,\Theta_i=3.16\}$,
		B: $\{V_0=6.51\times10^{-10}\,, \xi\sigma_f=0.6,\,\sigma_f=0.0258,\,\sigma_i=10.47,\,\Theta_i=0.079\}$ and 
		C: $\{V_0=9.78\times10^{-10}\,, \xi\sigma_f=0.6,\,\sigma_f=0.0111,\,\sigma_i=9.86,\,\Theta_i=0.079\}$. For all three cases we fix $C_\sigma=0.1,\,$ $m_0=0.5$ and $C_\Theta=0.01$. }
	\end{figure}

Now, as expected in inflationary backgrounds, the potential of the clock field in Eq.~\eqref{eq:potI} receives a correction of the form $m_0^2\sigma^2$. Here too, we plot some representative examples in Fig.~\ref{fig:BGModelII}. The clock field $\sigma$ first rolls down the quadratic potential. The region around the minimum, however, is dominated by the contribution of the $\propto \,\left[1-\exp\left(-\sigma^2/\sigma_f^2\right)\right]$ term in the potential and  $\sigma$ decelerates before it finally undergoes damped oscillations around $\sigma=0$. As in the first model, the second stage of inflation is then driven by $\Theta$. This richer dynamics produces a double bump in the first slow-roll parameter $\epsilon$, as can be seen from the right panel of Fig.~\ref{fig:BGModelII}. The first broader bump is caused by the change in the curvature of the potential that decelerates the motion of $\sigma$, whereas the second is the sharp feature in the $\epsilon$ parameter produced by the oscillations of the clock field, which was also seen in Model I. Differently from Model I, even though slow-roll is temporarily violated, here the $\epsilon$ parameter never becomes larger than 1 and the Universe never stops inflating.

\begin{figure}
		\begin{center} 
			\includegraphics[width=\columnwidth]{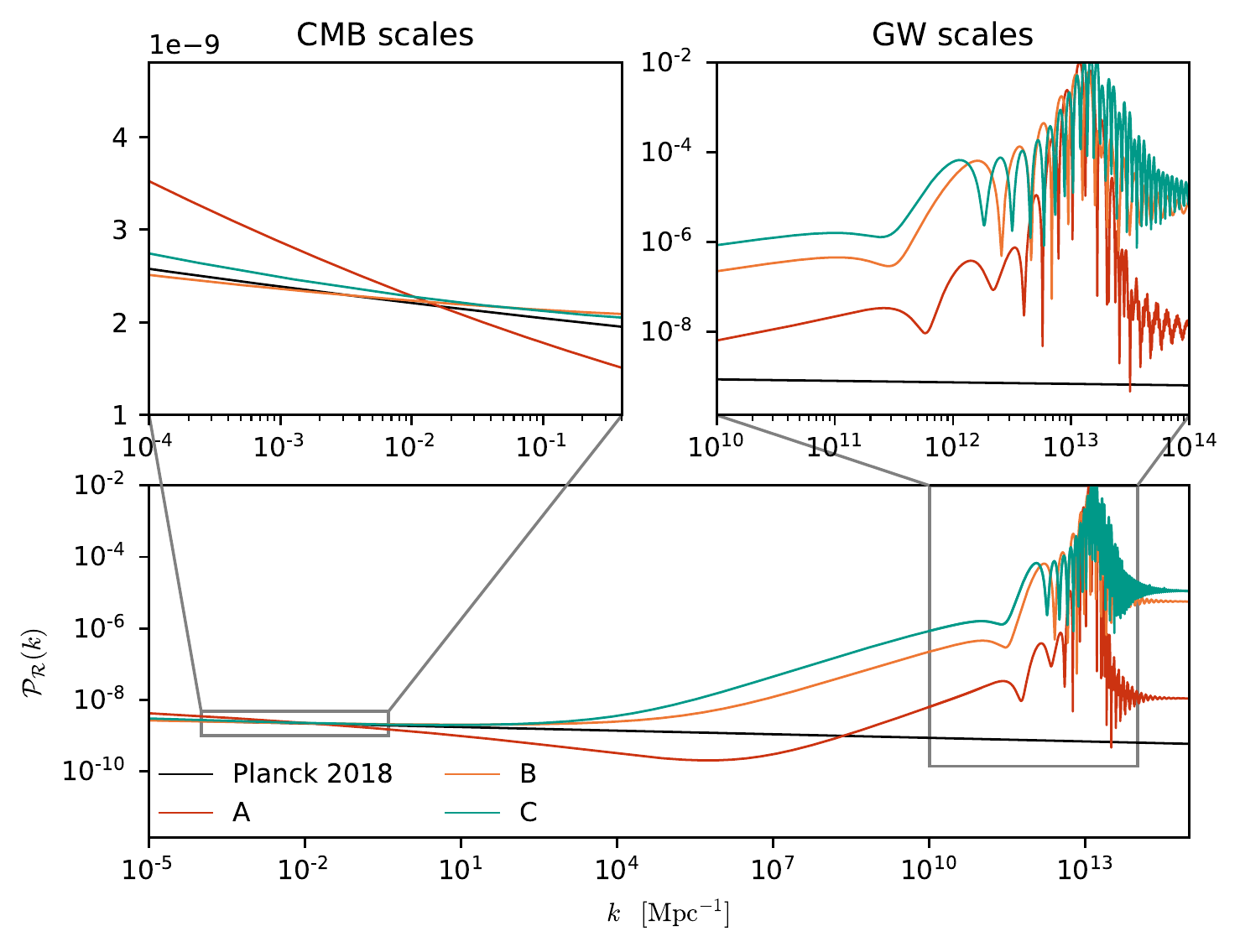}	
		\end{center}
		\caption{\label{fig:PkModelII} PPS [bottom] zoomed at CMB [top-left] and LISA [top-right] scales  for Model II. The parameters used in the three examples are the one reported in the caption of Fig.~\ref{fig:BGModelII}. For a comparison, we also plot the Planck 2018 power-law spectrum with $n_s=0.9665$  and $A_s=2.09\times 10^{-9}$ at the pivot scale $k=0.05\, {\rm Mpc}^{-1}$ \cite{Akrami:2018odb}. }
	\end{figure}

As discussed in Ref.~\cite{Chen:2014cwa}, the effect of the mass term for $\sigma$ is to suppress the PPS at the scales larger than the ones that cross the horizon during the clock field oscillations. This can be seen from Fig.~\ref{fig:PkModelII}, where we plot the PPS for a large range of scales to better appreciate this suppression. Since the power at large scales, that has to be matched to the COBE normalization, is now suppressed, the small scale plateau is now enhanced compared to Model I (see Fig.~\ref{fig:PkModelI}) and we need a smaller kinetic coupling compared to Model I.  As we will see,  the larger amplitude of the plateau at small scales will have interesting consequences when discussing the detectability of the induced SGWB.

For the examples in Fig.~\ref{fig:PkModelII}, $n_s=0.9011,\, 0.9754,\, 0.9612$, showing that case A is ruled out by Planck, but B and C are in complete agreement with the Planck  results. In fact, as for Model I, the model parameters can always be adjusted to make $n_s$ compatible with observations. 

Unfortunately, the specific model under consideration leads to an overproduction of Gravitational Waves at CMB scales and the  tensor to scalar ratio $r$ of the order of $\mathcal{O}(0.1)$ is not consistent with observations \cite{Akrami:2018odb}. This is quite natural,  being $\epsilon$  large during the first stage of inflation,  similarly to chaotic inflationary models.
A way out to get a lower tensor to scalar ratio, and at the same time retain the same dynamics generating the spectral features of the PPS in Fig.~\ref{fig:PkModelII}, could be to  consider a potential for $\sigma$ with a nearly inflection point that makes $\sigma$ decelerate as in Fig.~\ref{fig:BGModelII} \cite{Garcia-Bellido:2017mdw,Ballesteros:2017fsr,Dalianis:2018frf,Dalianis:2020cla,Ballesteros:2020qam}.
We defer the study of a model producing the dynamics studied in this Section, while satisfying all the observational data, to future works.
The purpose of this paper is to classify the types of features at small scales that can be obtained in this model.

%%%%%%%%%%%%%%%%%%%%%%%%%%%%%%%%%%%%%%%%%%%%%%%%%%%%%%%%%%%%%%%%%%%%%%%%%%%%%%%

\subsection{Model III}
%\section{Model III: large field inflation in the $\Theta$ direction}
\label{sec:ModelIII}	
As a third example, we consider the following modification by changing the shape of potential in the $\Theta$ direction, again introduced in Refs.~\cite{Chen:2014cwa}:
\begin{equation}
\label{eq:potIII}
U(\Theta)=V_{\rm inf}\frac{\left(m_\Theta\sqrt{V_0}\right)^2}{2}\Theta^2,\quad
	V(\sigma)=V_{\rm inf}\,C_\sigma\,\left[1-\exp\left(-\sigma^2/\sigma_f^2\right)\right].
\end{equation}
The potential in the $\Theta$ direction now acquires the form of chaotic inflation, therefore inflation ends smoothly without the need for an extra hybrid field to assist it. A potential very similar Eq.~\eqref{eq:potIII} was also studied in Ref.~\cite{Braglia:2020eai}  where two of us presented results similar to the ones shown in this Subsection.

\begin{figure}
		\begin{center} 
			\resizebox{214pt}{172pt}{\includegraphics{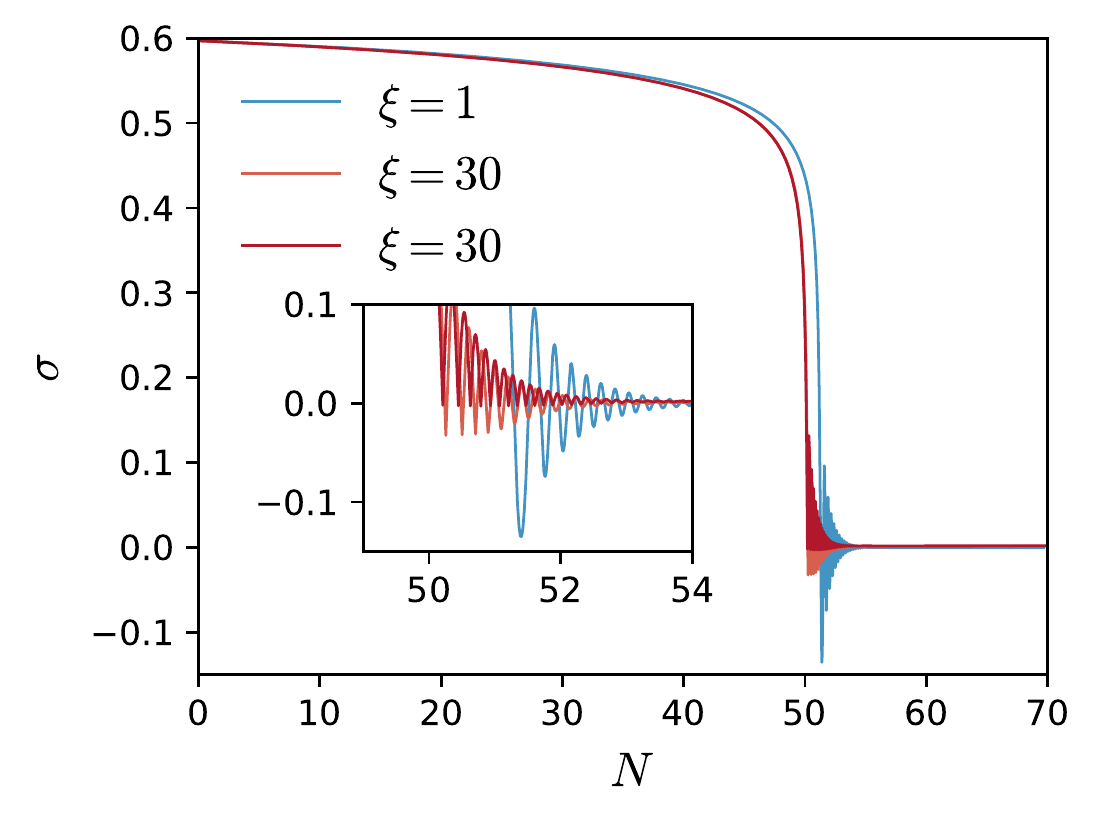}}
		\resizebox{214pt}{172pt}{\includegraphics{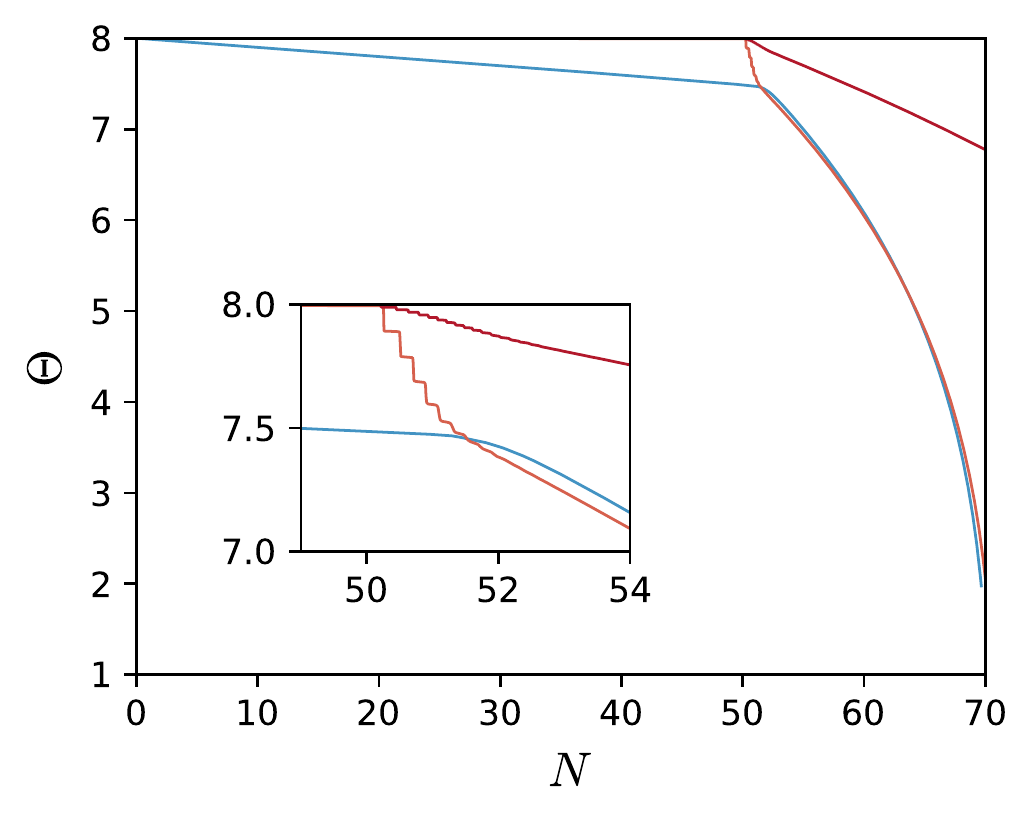}}
		\end{center}
		\caption{\label{fig:BGModelIII} Evolution of $\sigma$ [left] and $\Theta$ [right] for Model III.  The parameters used in the three examples are 
		$\{C_\sigma=500\,, \sigma_f=0.18,\,\sigma_i=3.27,\,\Theta_i=8\}$ and we vary $\xi\in [1,\,30,\,300]$.  }
	\end{figure}

In Fig.~\ref{fig:BGModelIII}, we show an example of the background dynamics in our model. As can be seen the dynamics of the two scalar field is quite peculiar. As in the previous Sections, the clock field eventually starts to oscillate. However, given the large field potential for $\Theta$, now the interaction between the two scalar fields are stronger and for a large coupling $\xi$ the zero point of the clock field oscillations changes. For very large $\xi$, as can be appreciated from the inserts in Fig.~\ref{fig:BGModelIII}, we arrive at the limiting situation in which $\sigma=0$ becomes a barrier for $\sigma$.

\begin{figure}
		\begin{center} 
		\resizebox{214pt}{172pt}{\includegraphics{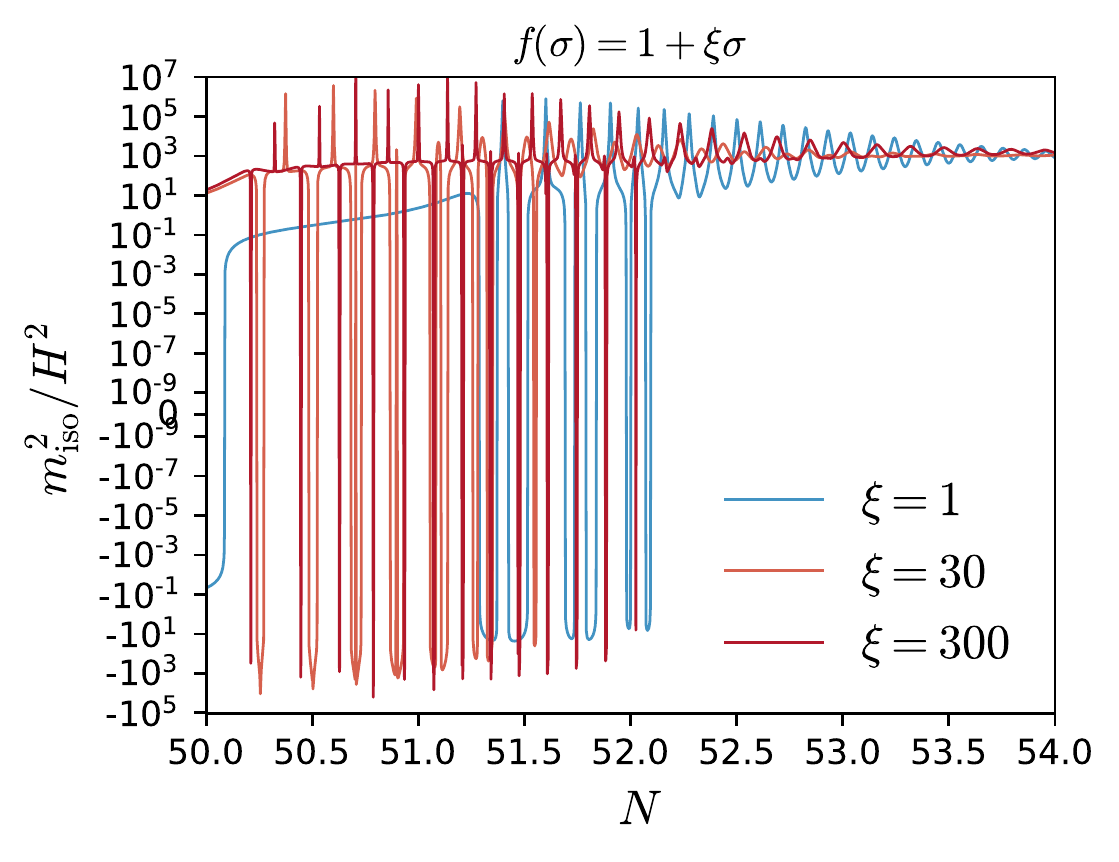}}
		\resizebox{214pt}{172pt}{\includegraphics{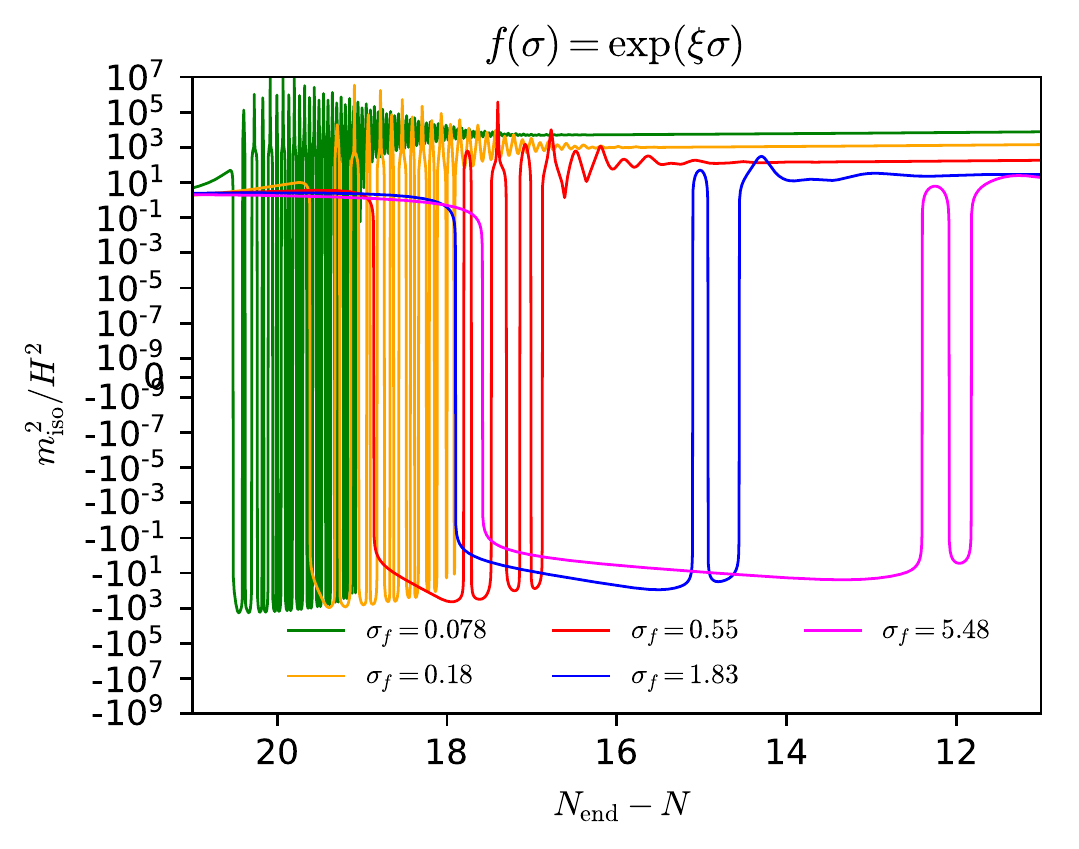}}
		\end{center}
		\caption{\label{fig:MisoModelIII} Evolution of the  mass squared of isocuvature perturbations  defined in Eq.~\eqref{eq:effmassiso} around the time when slow-roll is violated for the coupling $f(\sigma)=1+\xi\sigma$ [left] and for the coupling $f(\sigma)=\exp(\xi\sigma)$ [right]. The parameters used in the former case are the ones used in Fig.~\ref{fig:BGModelIII}, while in the second case we fix $C_\sigma=500$, $m_\Theta=1$ and use the following sets for the other parameters $\{V_0=2.14\times10^{-12},\,\xi=7.37,\,\sigma_f=5.48,\sigma_i=9.45,\,\Theta_i=4.4\},\,$  $\{V_0=1.1\times10^{-13},\,\xi=8.85,\,\sigma_f=1.83,\sigma_i=4.29,\,\Theta_i=6.5\},\,$   $\{V_0=5.5\times10^{-15},\,\xi=12.5,\,\sigma_f=0.55,\sigma_i=1.57,\,\Theta_i=8.05\},\,$
		   $\{V_0=7.0\times10^{-17},\,\xi=18.9,\,\sigma_f=0.078,\sigma_i=0.276,\,\Theta_i=7.8\},\,$.}
	\end{figure}
 In order to explain why it is not possible to obtain a substantial amplification of curvature perturbations within this setting, we plot the squared mass of isocurvature modes, defined in Eq.~\eqref{eq:effmassiso}, in the left panel of Fig.~\ref{fig:MisoModelIII} for some very large values of $\xi$.
As discussed above, the mechanism to amplify curvature perturbations that we are studying in this paper relies on the transient tachyonic growth of isocurvature modes \cite{Braglia:2020eai}. As can be seen, although $m_{\rm iso}^2$ does become negative, it does so for such a small amount of time, that it is not possible for the isocurvature perturbations to grow enough and source a sizeable amplification of the curvature ones. Setting a larger $\xi$ only shrinks the width of the oscillations in $m_{\rm iso}^2$ therefore makes it impossible to obtain any bump in the PPS.

			\begin{figure}
		\begin{center} 
					\includegraphics[width=\columnwidth]{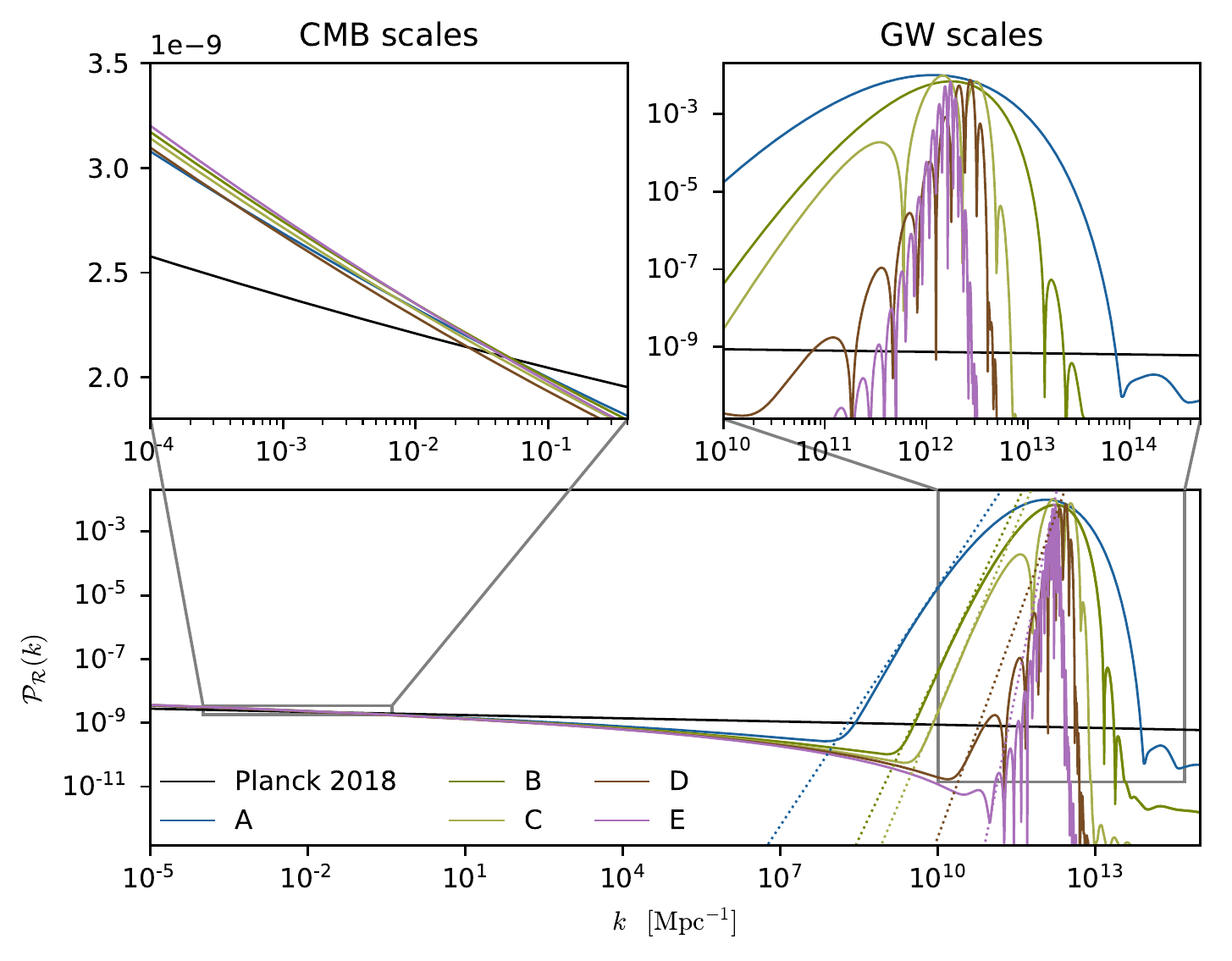}	
		\end{center}
		\caption{\label{fig:PkModelIII} PPS [bottom] zoomed at CMB [top-left] and LISA [top-right] scales  for Model III.  We fix $m_\Theta=1$ and use the following parameters:
		A: $\{V_0=2.3\times10^{-12},\,C_\sigma=500,\,\xi=7.5,\,\sigma_f=5.477,\sigma_i=9.311,\,\Theta_i=4.4\},\,$
B: $\{V_0=1.34\times10^{-13},\,C_\sigma=500,\,\xi=9.0,\,\sigma_f=1.826,\sigma_i=4.290,\,\Theta_i=6.7\},\,$
C: $\{V_0=6.3\times10^{-15},\,C_\sigma=550,\,\xi=12.2,\,\sigma_f=0.574,\sigma_i=1.649,\,\Theta_i=8.25\},\,$
D: $\{V_0=5.0\times10^{-16},\,C_\sigma=500,\,\xi=14.5,\,\sigma_f=0.183,\sigma_i=0.597,\,\Theta_i=8.0\},\,$
E: $\{V_0=8.2\times10^{-17},\,C_\sigma=500,\,\xi=18.9,\,\sigma_f=0.078,\sigma_i=0.277,\,\Theta_i=8.0\},\,$. For a comparison, we also plot the Planck 2018 power-law spectrum with $n_s=0.9665$  and $A_s=2.09\times 10^{-9}$ at the pivot scale $k=0.05\, {\rm Mpc}^{-1}$ \cite{Akrami:2018odb}.		}
	\end{figure}

From the point of view model building, a solution is to change the non-canonical coupling adopting another functional form, i.e. $f(\sigma)=\exp(\xi\sigma)$, as in Refs.~\cite{Braglia:2020eai,Braglia:2020fms}. We show the resulting  mass squared of isocurvature modes for different choices of parameters in the right panel of Fig.~\ref{fig:MisoModelIII}. Now $m_{\rm iso}^2$ can become negative for a larger amount of  time so that  some of the isocurvature modes that cross the horizon during that time can grow and sizeably amplify curvature amplification. Note that, as the effective mass of the massive field increases, i.e. $\sigma_f$ decreases, $m_{\rm iso}^2$ shows narrower peaks with a higher frequency. This has mainly two consequences. First, we expect a series of bumps (as in model I and II) for the modes that cross the horizon during slow-roll violation. For small $\sigma_f$ some of the isocurvature modes cross the horizon when $m_{\rm iso}^2<0$ and gets amplified and some cross it when $m_{\rm iso}^2>0$ and get suppressed.  The second is that, having less time to grow, we need a stronger coupling $\xi$ for models with a smaller $\sigma_f$ to reach the same peak amplitude.

Assuming the exponential coupling, we plot in Fig.~\ref{fig:PkModelIII} the PPS for some examples. As in Model I, we observe a series of bumps with the character of a sharp feature signal. However, due to the large field behavior of the second field $\Theta$, which sizeably contributes to the total energy density, we are forced here to choose a large potential dip, i.e.~$C_\sigma\gg 1$ in order to enhance the clock signal. Because of that, only the sharp feature signal is amplified and after the peak with the largest amplitude the PPS rapidly decays and the resonant feature is suppressed. 

Interestingly, in the limiting situation in which the effective mass of $\sigma$ is very small, i.e. $\sigma_f$ is very large, only the first bump in the sinusoidal running of the sharp signal is amplified. This results in a very broad peak structure of the bump, as analyzed in Ref.~\cite{Braglia:2020eai}, possibly leading to the formation of PBH with a broad mass spectrum if its amplitude is high enough (see also Refs.~\cite{Palma:2020ejf,Fumagalli:2020adf}). Other multifield inflationary models that can produce a large bump in the PPS, although with a shape different from the one studied here, were studied in Refs.~\cite{Pi:2017gih,Aldabergenov:2020bpt,Nanopoulos:2020nnh,Aldabergenov:2020yok,Gundhi:2020kzm,Gundhi:2020zvb,Garcia:2020mwi}

As for CMB scales, instead, we always obtain  a spectral index redder than Planck constraints. Specifically, we get $n_s=
0.9397,\,
0.9356,\,
0.9351,\,
0.9369,\,
0.9327$ for the model A, B, C, D and E respectively. This was observed also in Ref.~\cite{Braglia:2020eai} and points to the need of a different potential for the field $\sigma$, which can restore the concordance with Planck results.

Also note that the rise of the power spectrum can be much faster than $n_s-1=4$, which is the limit in single-field inflationary models \cite{Byrnes:2018txb,Carrilho:2019oqg}, with interesting phenomenological implications \cite{Gow:2020bzo,Unal:2020mts}. In particular, we get that the spectrum at small scales grows with a spectral index of approximately $n_s=
3.5,\,
4.5,\,
4.9,\,
5.5,\,
9.1$ for the model A, B, C, D and E respectively. The fact that the growth of the PPS in multifield model can easily overcome the single-field limit on $n_s$ was also demonstrated in Refs.~\cite{Palma:2020ejf,Fumagalli:2020adf,Braglia:2020eai}.

\section{Generated SGWB and detectability with future space-based interferometers}
\label{sec:SGWBandDetect}

The large scalar perturbations presented in the previous Section inevitably act as a source of gravitational waves when they re-enter the horizon during radiation era \cite{Matarrese:1993zf,Matarrese:1997ay,Acquaviva:2002ud,Carbone:2004iv,Ananda:2006af,Baumann:2007zm,Assadullahi:2009nf}.
The expression for the energy density of the gravitational waves  is~\cite{Baumann:2007zm,Espinosa:2018eve}:
	\begin{align}
	\label{eq:GW}
	\Omega_{\textup{GW}}=&\frac{\Omega_{r,0}}{36}\int_{0}^{\frac{1}{\sqrt{3}}}d\textup{d}\int_{\frac{1}{\sqrt{3}}}^{\infty}ds\,\left[\frac{(d^2-1/3)(s^2-1/3)}{s^2-d^2}\right]^2\notag\\
	&\cdot\mathcal{P}_{\mathcal{R}}\left(\frac{k\sqrt{3}}{2}(s+d)\right)
	\mathcal{P}_{\mathcal{R}}\left(\frac{k\sqrt{3}}{2}(s-d)\right)[\mathcal{I}_c(d,s)^2+\mathcal{I}_s(d,s)^2].\end{align}
	In the equation above, $\Omega_{r,0}\simeq 8.6\times10^{-5}$ is the density of radiation today and  the functions in the integral are explicitly given by
	\begin{equation}
	\begin{aligned}
		{\cal I}_c(x,y) &=4  \int_0^\infty \dd\tau \, \tau (-\sin \tau)  \Big[ 2T(x\tau)T(y\tau) + \Big(T(x\tau)
		+ x\tau\, T'(x\tau) \Big)\Big(T(y\tau) + y\tau\, T'(y\tau) \Big) \Big],
		\\
		{\cal I}s(x,y) &= 4 \int_0^\infty \dd\tau \, \tau (\cos \tau)  \Big\{ 2T(x\tau)T(y\tau) + \Big[T(x\tau) + x\tau\, T'(x\tau) \Big] \Big[T(y\tau) + y\tau\, T'(y\tau) \Big] \Big\},
	\end{aligned}	\label{eq: Ic, Is}
\end{equation}
where $T(k \eta)$ is the standard transfer function in the radiation dominated era  \begin{equation}
T(z)= \frac{9}{z^2}\left[ \frac{\sin (z/\sqrt 3)}{z/\sqrt 3} -\cos(z/\sqrt 3) \right].
\end{equation}
	A complete analytical expression of the functions $\mathcal{I}_{c,s}$ is given in Eqs.~(D.1) and (D.2) of Ref.~\cite{Espinosa:2018eve} (see also Ref.~\cite{Kohri:2018awv}). Note that an implicit assumption that is made in the literature in deriving Eq.~\eqref{eq:GW} is a $\Lambda$CDM evolution. In fact, the induced SGWB can be used as a tool to test the thermal history of the Universe, as additional energetic contents in the Universe with an equation of state different from $w=1/3$ can modify the equation above \cite{Domenech:2019quo,Domenech:2020kqm}. For simplicity, however, we use the standard formula \eqref{eq:GW}.

%			\begin{figure}
%		\begin{center} 
%			\resizebox{214pt}{172pt}{\includegraphics{GW188mm.pdf}}	
%			\resizebox{214pt}{172pt}{\includegraphics{GW288mm.pdf}}	
%		\end{center}
%		\caption{\label{fig:OmegaGW} Features that can be produced in our models. In the left panel, we plot the  feature obtained using the spectrum labeled with A in Fig.~\ref{fig:PkModelIII} and the  feature obtained using  the spectrum labeled with C  Fig.~\ref{fig:PkModelIII}.  In the right panel, we plot the feature obtained using  the  spectrum labeled with C in Fig.~\ref{fig:PkModelI} and the  feature using the  spectrum labeled with C Fig.~\ref{fig:PkModelII}. The terminology used to label the spectra is introduced later in Section~\ref{sec:temp_est}.   We plot the sensitivity curves for LISA~\cite{Audley:2017drz,Caprini:2015zlo}  BBO and ultimate DECIGO~\cite{Yagi:2011wg} and Magis-AION-space~\cite{Coleman:2018ozp}.  }
%	\end{figure}

%In Fig.~\ref{fig:OmegaGW}, we plot the SGWB generated starting from some example cases presented in the previous Sections. As can be seen, the resulting $\Omega_{\rm GW}$ overlap the sensitivity curves of several planned GW space-based interferometers, suggesting that a GW signal from two-field inflation can be detected in the future. Moreover, the four cases plotted in Fig.~\ref{fig:OmegaGW} visibly show very distinct spectral shape. 

We now analyze the spectral features of the SGWB that can be produced in our model and discuss their detectability with future GW experiments. To do so, we follow two  independent and complementary approaches. The first is the approach of signal reconstruction, whereas in the second approach we derive templates for $\Omega_{\rm GW}(f)$ describing the spectral features produced by our models that can be used in future for data analysis.

			\begin{figure}
		\begin{center} 
			\resizebox{214pt}{172pt}{\includegraphics{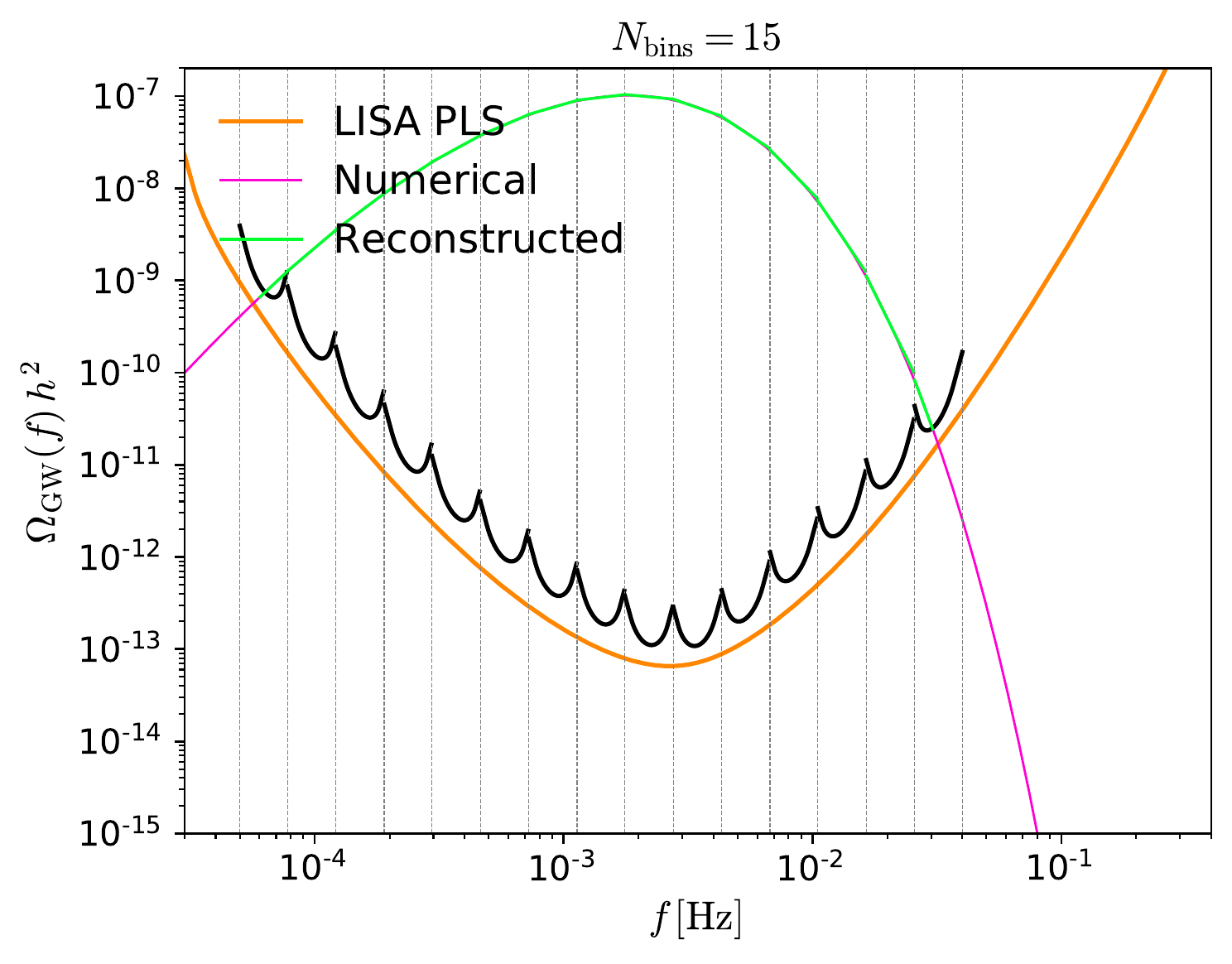}}
			\resizebox{214pt}{172pt}{\includegraphics{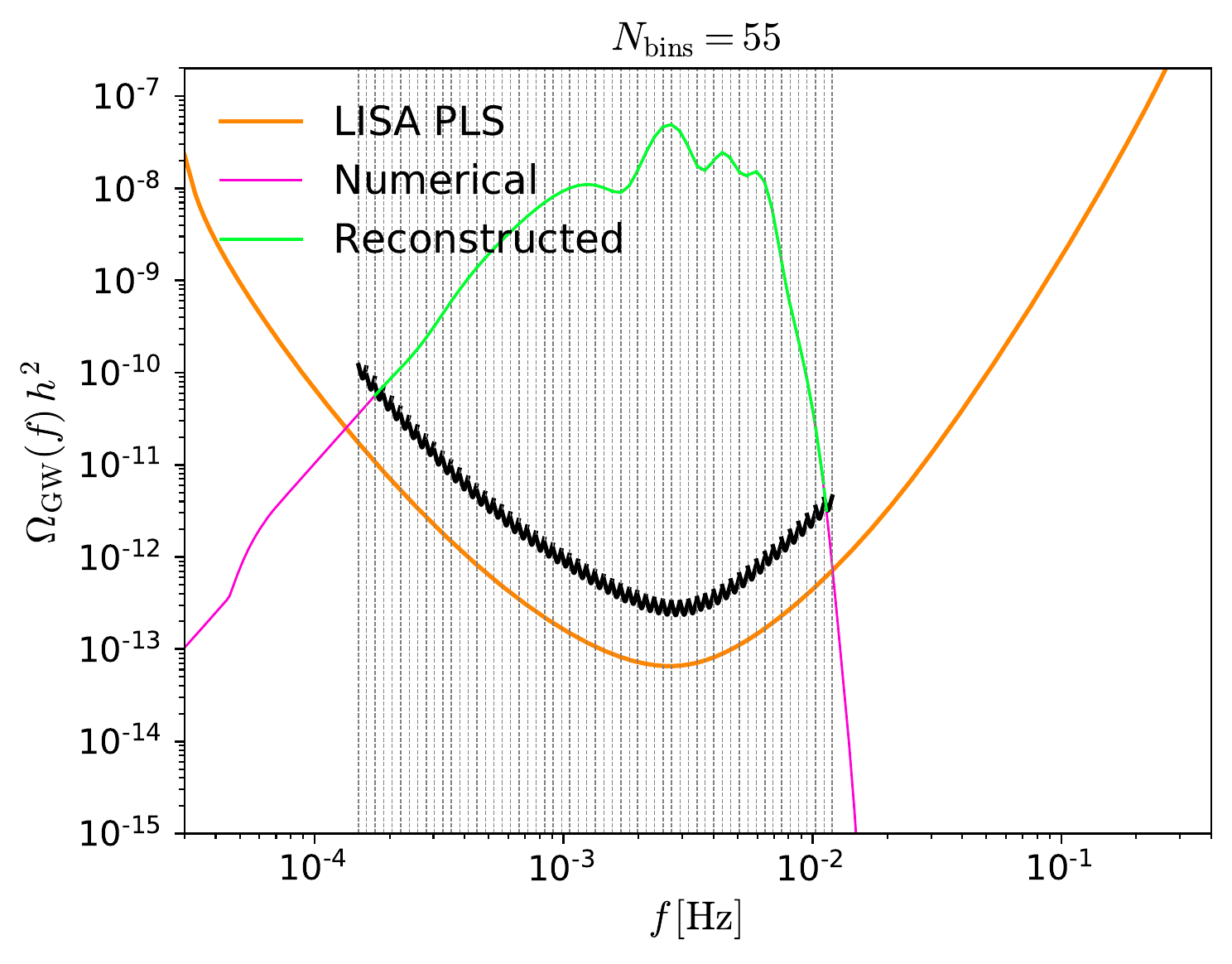}}
		\end{center}
		\caption{\label{fig:PLSI}  Reconstructed spectra for the bump [left] and sinusoidal [right] features. We also plot the numerical spectra in magenta lines and the LISA binned (unbinned) PLS in black (orange) lines.}
	\end{figure}

\subsection{The approach of signal reconstruction}
\label{sec:reconstruction}
We start by discussing the signal reconstruction approach. For simplicity, we will focus on SGWB spectra that fall within the sensitivity range of the future LISA mission \cite{Audley:2017drz}. Note however, that features at larger scales, probed by Pulsar Timing Array experiments, or smaller scales, probed by the next generation of space-based GW interferometers such as BBO and DECIGO \cite{Yagi:2011wg}, are also possible  if the massive field starts to oscillate earlier or later during inflation respectively. Moving the feature even closer to the end of inflation we enter in the regime of Einstein Telescope. Being the sensitivity to $\Omega_{\rm GW}$ better (worse) than the LISA one for BBO and DECIGO (Einstein Telescope), the results are more pessimistic (optimistic) than what  we would obtain considering such interferometers.

In the following, we adopt the formalism of the binned Power-Law integrated Sensitivity (PLS) curve introduced in Refs.~\cite{Caprini:2019pxz,Flauger:2020qyi}. Although a more quantitative analysis would require more advanced tools to reconstruct the signal more efficiently, such as the SGWBinner approach of Refs.~\cite{Caprini:2019pxz,Flauger:2020qyi}, the binned PLS procedure is very useful to show at which level the signal can be reconstructed by LISA and if it can be distinguished from a signal with a different spectral shape, at least qualitatively.

The procedure works as follows. The signal to noise SNR of the SGWB is given as
\begin{equation}
\label{eq:SNR}
    {\rm SNR}=\sqrt{T\int_{f_{\rm min}}^{f_{\rm max}}\,df\,\left(\frac{\Omega_{\rm GW} (f)}{\Omega_s(f)}\right)^2     }
\end{equation}
where $\Omega_s(f)$ is the energy density calculated from the noise Power Spectral Density (PSD)\footnote{We adopt the PSD for the LISA TDI X channel, see e.g. Ref.~\cite{Smith:2019wny}. }.
Within the original PLS method, one assumes that the SGWB is described by a power-law of the form $\Omega_{\rm GW}=A_* f^\alpha$ over the whole sensitivity range and computes the above equation to find the minimum value of $A_*$ that gives a target ${\rm SNR}_{\rm thr}$. Then the procedure is repeated for a range of negative and positive values of $\alpha$. The envelop of the largest values of $A_* f^\alpha$ at each frequency constitutes the PLS curve \cite{Thrane:2013oya}. In this paper, we compute the PLS for LISA using the schNell code\footnote{\href{https://github.com/damonge/schNell/branches}{https://github.com/damonge/schNell/branches}} introduced in Ref.~\cite{Alonso:2020rar} and consider a threshold value of ${\rm SNR}_{\rm thr}=10$, as suggested by previous studies \cite{Caprini:2019pxz}, and the nominal observational time of $T=4$ years.

As discussed above, however, the frequency dependence of the SGWB produced in our models is definitely not well represented by a power-law, or at least not over the whole sensitivity range of LISA. The most straightforward way to assess the power of LISA to reconstruct the spectral shape of a SGWB is to divide the sensitivity range in equi-log-spaced bins, such that $\Omega_{\rm GW}$ can be well represented by a power-law within each bin, and compute the PLS for each bin. As before, for each bin, the part of $\Omega_{\rm GW}$ that is larger then the  PLS curves can be detected with a signal to noise ratio ${\rm SNR}_{\rm thr}$. Therefore, in principle, any signal can be reconstructed by dividing the frequency range in many bins. However, as the number of bins grows, the sensitivity in the bins is substantially degraded and a large ${\rm SNR}$ is needed.

\begin{figure}
\begin{center} 
	\resizebox{214pt}{172pt}{\includegraphics{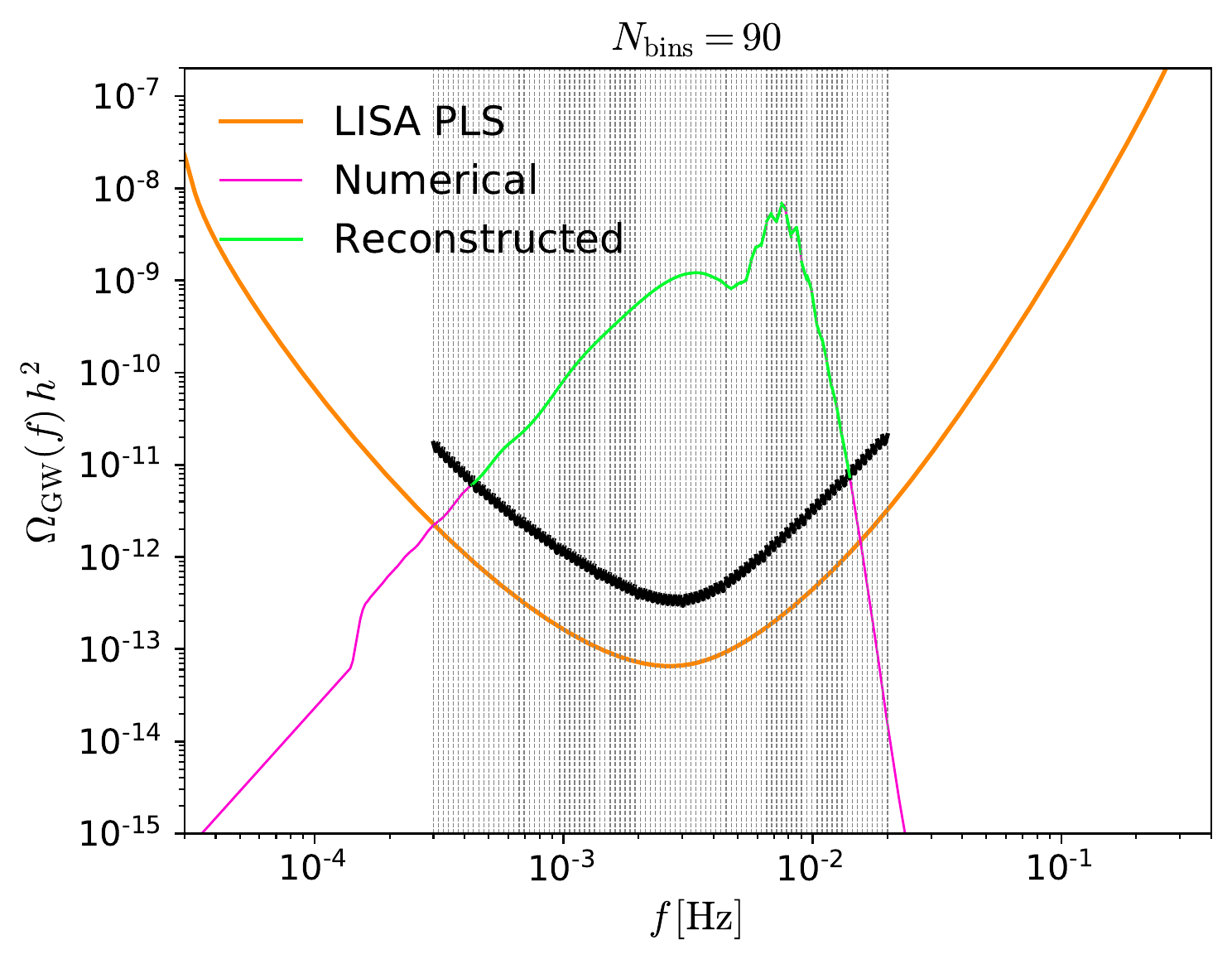}}
	\resizebox{214pt}{172pt}{\includegraphics{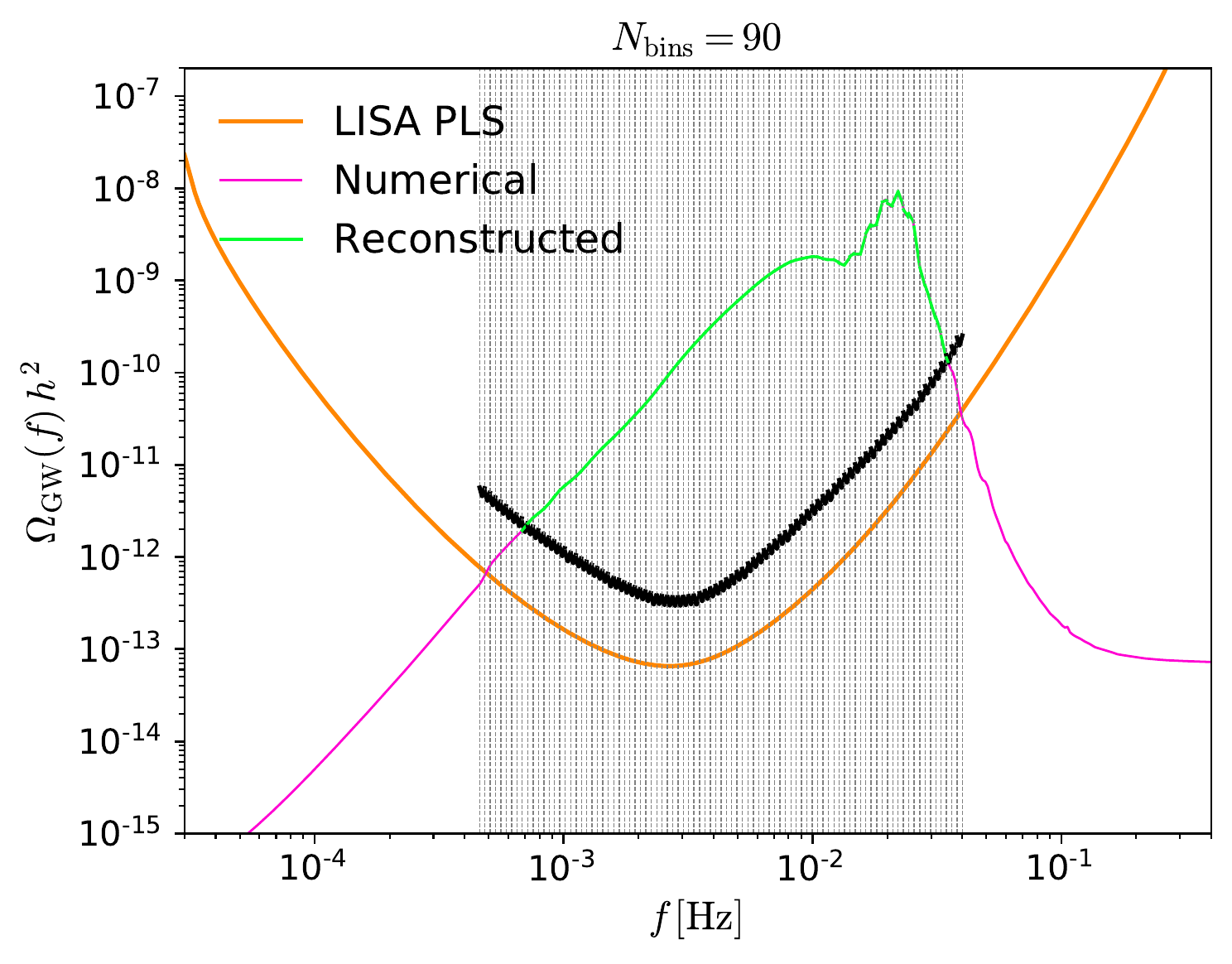}}
\end{center}
\caption{\label{fig:PLSII} Reconstructed spectra for resonant feature 1 [left] and 2 [right].  We also plot the numerical spectra in magenta lines and the LISA binned (unbinned) PLS in black (orange) lines.}
\end{figure}

We present our analysis in Figs.~\ref{fig:PLSI} and \ref{fig:PLSII} where we plot the LISA unbinned PLS, the binned PLS and some examples of numerical spectra that can be produced in our model. At this stage, we are only interested in discussing the reconstruction of these spectral features. Their nature and how they are generated is discussed in details in the next Subsection. For each bin within which the SGWB can be detected, we also plot the fitted power-law in blue.  The results of our analysis are summarized in Table~\ref{tab:bintable}, where we report the number of frequency bins and their width that are needed to reconstruct our spectra. Note that, in order not to overcrowd the plots, we only plot the bins within the frequency range where our signals are larger than the unbinned PLS. The number of bins is thus referred to that range. 

As evident, the signal that can be best reconstructed is the one in the left panel of Fig.~\ref{fig:PLSI}, that requires a bin width of $\Delta \log_{10}f =0.18$ and a number of bins as small as $N_{\rm bins}=15$ in order to be  correctly reconstructed. Note that our results obtained with the binned PLS are more pessimistic than those found in e.g. Refs.~\cite{Caprini:2019pxz,Flauger:2020qyi}, where a broken power-law and a log-normal $\Omega_{\rm GW}$, similar to our spectrum, were shown to be reconstructable with a smaller number of bins. With more advanced techniques such as those presented in Refs.~\cite{Caprini:2019pxz,Flauger:2020qyi}, some of the bins in Fig.~\ref{fig:PLSI} could indeed be merged without affecting the reconstruction at all. This shows the robusteness of the graphical method that we adopt. 

The signal in the right panel of Fig.~\ref{fig:PLSI}
 requires instead a smaller $\Delta \log_{10} f=0.034$ and a larger $N_{\rm bins}=55$, because of its more complicated spectral structure. Due to the degradation introduced in the procedure of binning the PLS, the signal reconstruction worsens at the extrema of the frequency range reported in Table~\ref{tab:bintable}, where the signal is smaller than the binned PLS. However, the most interesting part of the signal, where it shows some wiggles around its peaks, falls at the center of the unbinned LISA PLS, so it is not affected by this degradation and can be reconstructed.

A similar argument holds for the spectra shown in Fig.~\ref{fig:PLSII}, which show a very similar shape within the LISA frequency range. Now  the number of bins is as large as $N_{\rm bins}=90$ and their width is even smaller than the previous examples, i.e. $\Delta \log_{10} f=0.021$. Again, we note that the rising part of $\Omega_{\rm GW}$ at smaller frequency would not require such a large number of bins when using more advanced techniques \cite{Caprini:2019pxz,Flauger:2020qyi}. As in the previous example, the spectra cannot be reconstructed  at the extrema of the frequency range in Table~\ref{tab:bintable}.

\begin{table}[]
\begin{center}
\begin{tabular}{|l|l|l|l|}
\hline
SGWB signal type & $\Delta \log_{10} f$ & $N_\mathrm{bins}$ & Frequency Range \\ \hline
Bump             & 0.18              &        15           & $[5\times10^{-5}, 0.04]$                \\ \hline
Sinusoidal          & 0.034            &      55             & $[1.5\times10^{-4},\,0.012]$                 \\ \hline
Resonance 1            &       0.020       &       90            &$[3.0\times10^{-4},\,0.02]$                  \\ \hline
Resonance 2          & 0.021           &    90               & $[4.6\times10^{-4},\,0.04]$               \\ \hline
\end{tabular}
\caption{\label{tab:bintable} The bin-width and number of bins required for reconstruction of different types of SGWB signals. The frequency range that we divide in bins, reported in the third column, is defined as the one within which our signal is larger than the unbinned LISA PLS. }
\end{center}
\end{table}			

\subsection{The approach of template estimators}
\label{sec:temp_est}
Another approach, which is complementary to the approach of signal reconstruction just presented, is the one of template estimators.
Such approach has been shown to be very effective in the CMB analyses of primordial features and non-Gaussianities.
We therefore go on to derive templates that capture the spectral features of $\Omega_{\rm GW}$ produced in our models, leaving the details of how to apply them in data analyses to a future study. 

The spectra presented in the previous Subsection present some very distinct spectral shapes. In order to propose accurate templates\footnote{We note that an exact analytical expression for the SGWB produced by a log-normal peak in the PPS has been derived in Ref.~\cite{Pi:2020otn}. However, as already mentioned before, the peak in the power spectrum produced in our models is different from a log-normal one, and the formulae derived in Ref.~\cite{Pi:2020otn} do not reproduce our results.}, and to understand how they arise in our models, as well as to make contact with the types of feature produced in inflationary models  reviewed in the Introduction and in Section~\ref{sec:ModelIntro}, we now divide our results in four classes and discuss them in turn.  
\begin{itemize}
    \item {\em Bump feature}. In certain situations, when the mass of the massive field (and the frequency of its oscillations) is sufficiently small, only the first bump in the sinusoidal running of the sharp feature is substantially amplified \cite{Braglia:2020eai}. This case is mainly observed in Model III, where the oscillations after the largest bump   quickly decay. An example of such feature is the blue line in Fig.~\ref{fig:PkModelIII}, 
which we also plot in the left panel of Fig.~\ref{fig:OmegaBump} for convenience.
    
    The fact that the signal is very smooth and rounded around its peak, makes it easy to analytically model its observationally relevant part  using the following template consisting in a power-law, to model the larger scales, and  a log-normal peak with an exponential UV cutoff:

	\begin{equation}
	\label{temp:bumpy}
	h^2\Omega_{\rm GW}(f)= \begin{cases}
	A_1\,\left(\frac{f}{f_1}\right)^\alpha &\text{for $f\leq f_1$}\\
	A_3\,\exp\left[-d_1\ln\frac{f}{f_p}-d_2\ln^2\frac{f}{f_p}+d_4\frac{f}{f_p} \right]  &\text{for $f> f_1$}.
	\end{cases}
	\end{equation}
	As shown in Fig.~\ref{fig:OmegaBump}, the bump feature can be reproduced quite well by using the set of parameters given in Appendix~\ref{app:templates}. 
	
	The fact that the bump feature is nearly flat around its peak makes it a perfect candidate to explain the recent results by the North American Nanohertz Observatory for Gravitational waves (NANOGrav), that has recently reported evidence of a stochastic common-spectrum process \cite{Arzoumanian:2020vkk}. The NANOGrav result  can be interpreted as a SGWB signal fitted by a power-law with amplitude  at  $\Omega_{\rm GW}(5.5\, {\rm nHz}) \in[3 \times 10^{-10},\,2 \times 10^{-9}]$ and a spectral index $\alpha\in[-1.5,\,0.5]$ at 2$\sigma$ \cite{Arzoumanian:2020vkk}. In the context of our model, this would imply a massive field that starts to oscillate earlier during inflation. The possibility to fit  the NANOGrav result with the model introduced in Ref.~\cite{Braglia:2020eai}, where results similar to those presented were found, has been recently studied in Ref.~\cite{Vaskonen:2020lbd} and the analysis with a pure log-normal or exactly flat peak in the PPS has been performed in Refs.~\cite{Kohri:2020qqd} and \cite{DeLuca:2020agl} respectively, showing similar results.

			\begin{figure}
		\begin{center} 
			\resizebox{214pt}{172pt}{\includegraphics{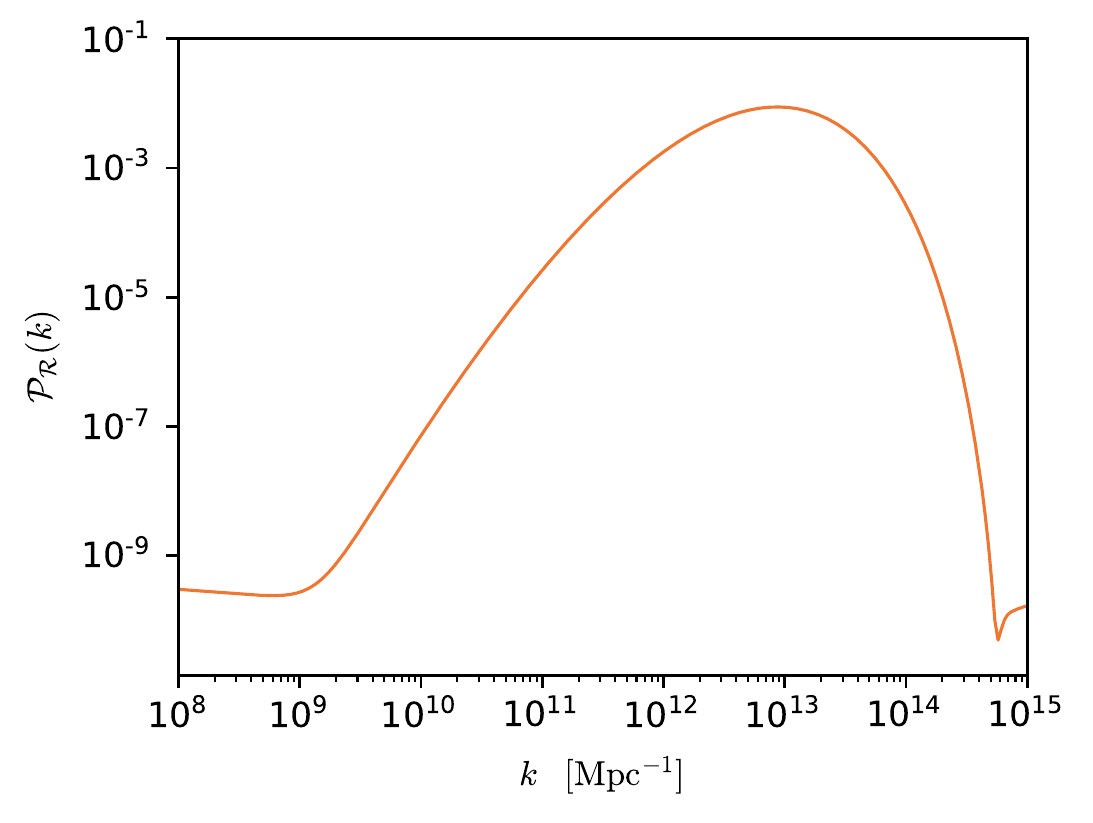}}	
			\resizebox{214pt}{172pt}{\includegraphics{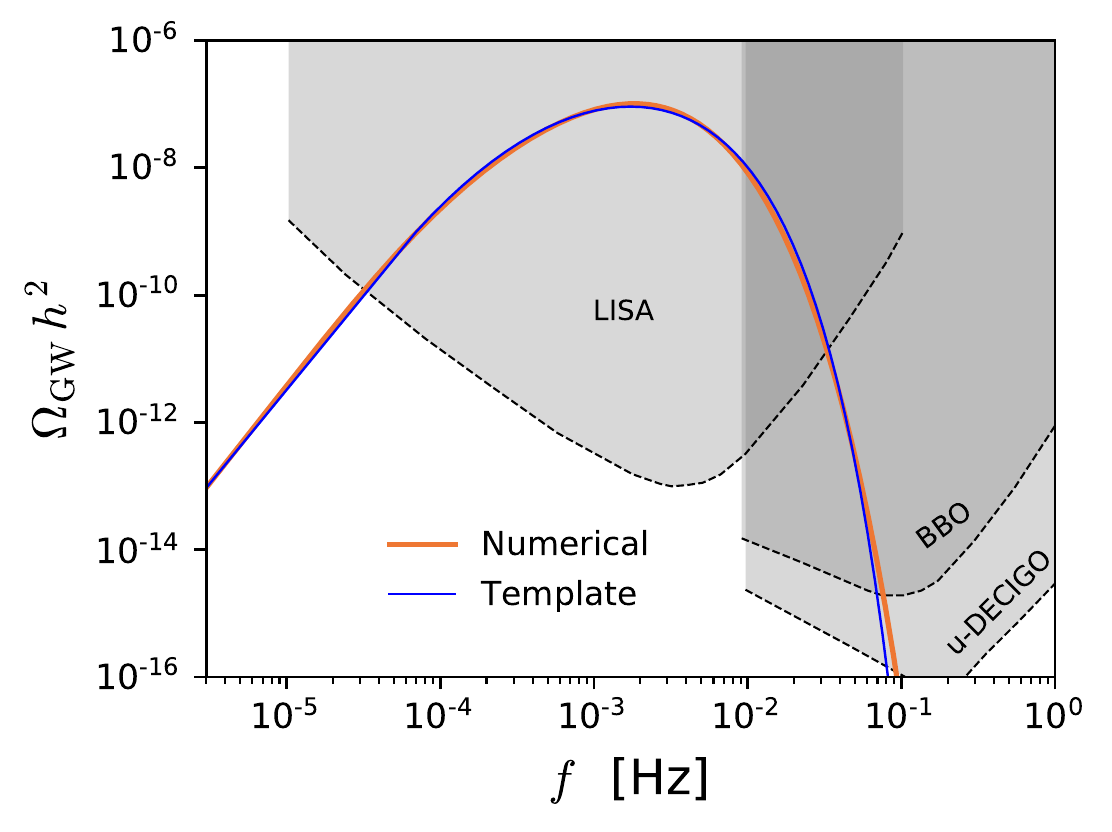}}	
		\end{center}
		\caption{\label{fig:OmegaBump} Primordial scalar power spectrum [left] and SGWB energy density [right] for the bump feature produced in Model III (blue line in Fig.~\ref{fig:PkModelIII}). In the right panel we plot the numerical spectrum (orange line) and the fit with the analytical template in Eq.~\eqref{temp:bumpy}.    We also plot the sensitivity curves for LISA~\cite{Audley:2017drz,Caprini:2015zlo},  BBO and ultimate DECIGO~\cite{Yagi:2011wg}.  }
	\end{figure}

    \item {\em Sinusoidal feature}. This class of features is produced in intermediate situations in which the effective  mass of the massive field is not large enough for the field to oscillate with a very high frequency, so that the spectrum present a series of quite broad peaks. As opposed to the bump feature, the sinusoidal one is produced by a mixture of sharp feature and clock signal, see also discussion in Sections \ref{sec:ModelI},~ \ref{sec:ModelII} and \ref{sec:ModelIII}. The light green line shown in Fig.~\ref{fig:PkModelII} of Model III, also reported in Fig.~\ref{fig:OmegaWiggly}, is an example of such a feature. 
    Another example of sinusoidal feature could be the spectrum labeled with A in Fig.~\ref{fig:PkModelI}. The signal presents a series of secondary peaks superimposed to  the broad peak structure $\Omega_{\rm GW}$. This suggests that the spectral shape can be modeled analytically starting from the bumpy feature template and adding some narrower log-normal peaks.
    
        Indeed, the observationally relevant part of the SGWB spectrum is captured by the following template consisting in the addition of a series of log normal functions to model the secondary peaks to the bumpy feature template in Eq.~\eqref{temp:bumpy}:

	\begin{equation}
\label{temp:wiggly}
h^2\Omega_{\rm GW}(x=f/f_p)= \begin{cases}
A_1\,\left(\frac{f}{f_1}\right)^{\alpha}&\text{for $f\leq f_1 $}\\
A_2\,\exp\left[-b_1(-\ln x -B_1)^\beta\right]&\text{for $f_1<f<f_2$}\\
A_3\Biggl\{\,\exp\left[-d_1\ln x\,-d_2\left(\ln x - D_2\right)^{\delta_2}\right]+&\\
A_4\exp\left[-g_1\ln x\,-\sum_{i=2}^3 g_i\left(\ln x - G_i\right)^{\gamma_i}\right]g(f)\Biggr\}&\text{for $f>f_2$}.
\end{cases}
\end{equation}
where $f_p$ is the frequency of the first peak and $A_2$ and $A_3$ are found by matching the templates at $f=f_1$ and $f=f_2$ and $g(f)=\left[1+l\sin \left(\omega\, (f-f_p) + \phi\right)\right]$.
The sinusoidal dependence on frequency is inherited from the sinusoidal dependence of the feature on scale, and is a signature of presence of a sharp feature.

As shown in  Fig.~\ref{fig:OmegaWiggly}, the signal can be reproduced quite well by using the set of parameters given in Appendix~\ref{app:templates}.

			\begin{figure}
		\begin{center} 
			\resizebox{214pt}{172pt}{\includegraphics{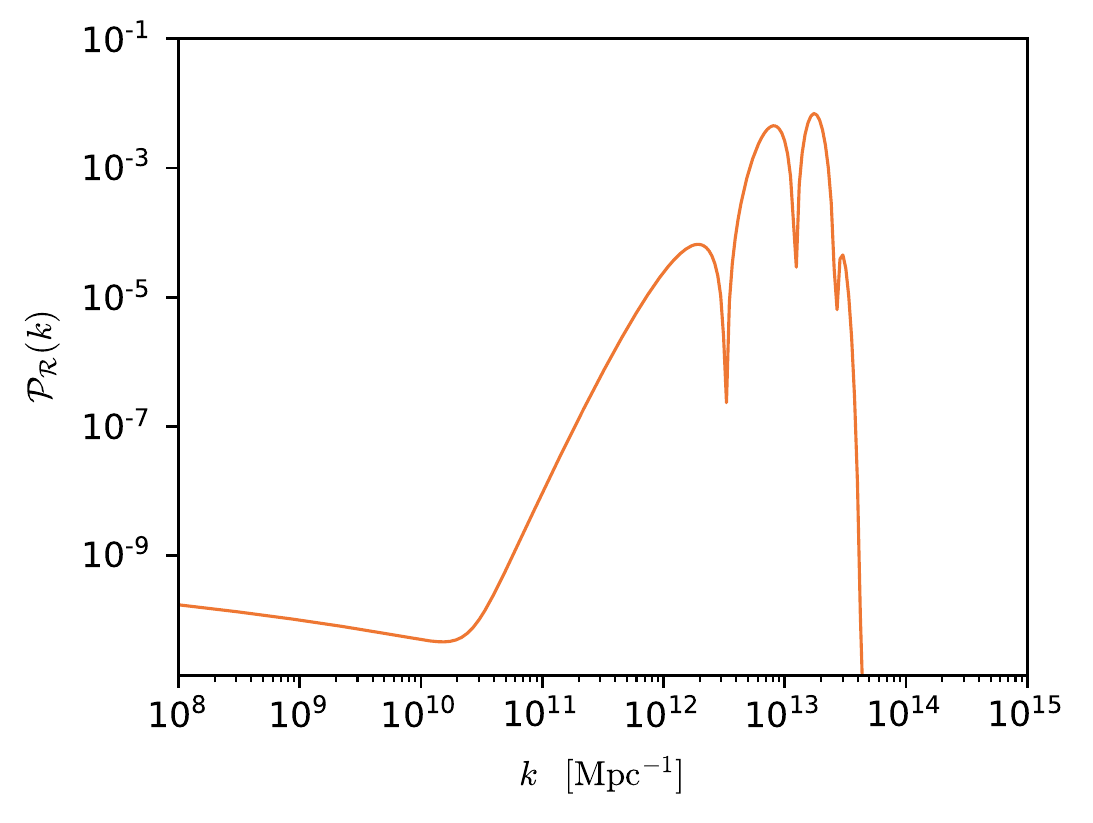}}
			\resizebox{214pt}{172pt}{\includegraphics{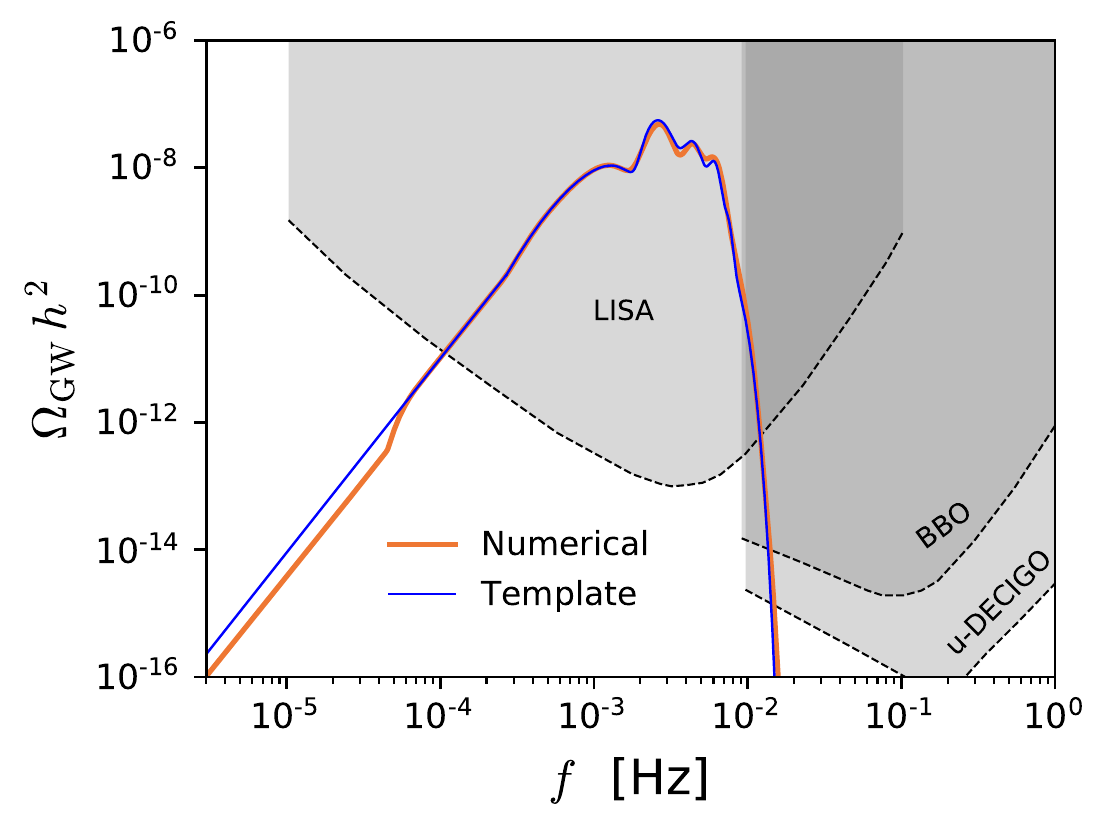}}	
		\end{center}
		\caption{\label{fig:OmegaWiggly} Primordial scalar power spectrum [left] and SGWB energy density [right] for the sinusoidal feature  produced in Model III (light green line in Fig.~\ref{fig:PkModelIII}). In the right panel we plot the numerical spectrum (orange line) and the fit with the analytical template in Eq.~\eqref{temp:wiggly}.    We also plot the sensitivity curves for LISA~\cite{Audley:2017drz,Caprini:2015zlo},  BBO and ultimate DECIGO~\cite{Yagi:2011wg}.  }
	\end{figure}

			\begin{figure}
		\begin{center} 
			\resizebox{214pt}{172pt}{\includegraphics{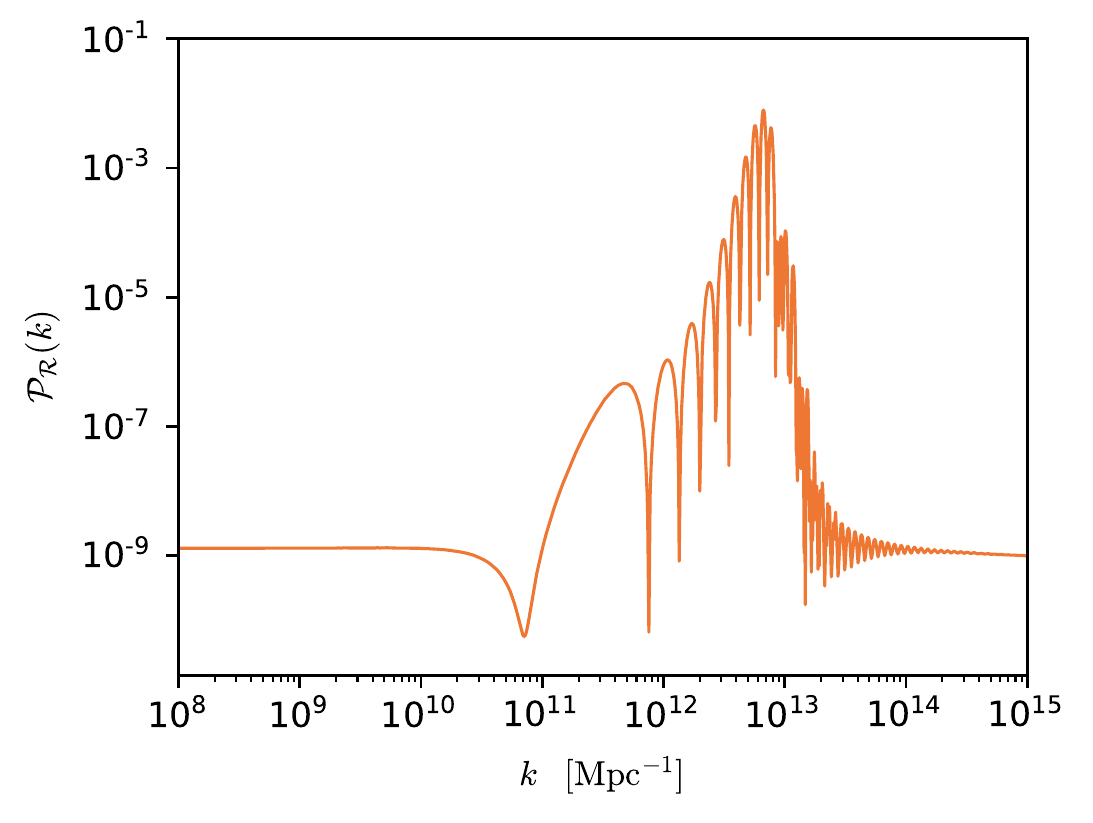}}
			\resizebox{214pt}{172pt}{\includegraphics{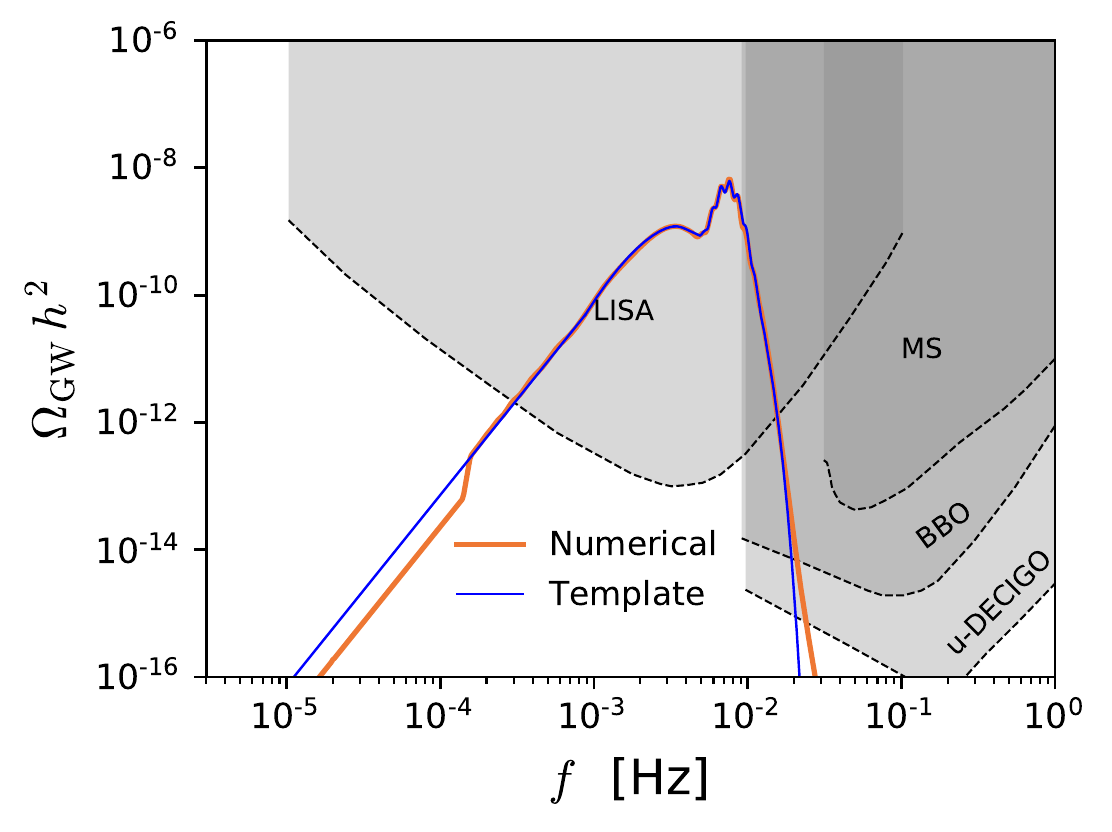}}	
		\end{center}
		\caption{\label{fig:OmegaDelta} Primordial scalar power spectrum [left] and SGWB energy density [right] for the resonant feature 1  produced in Model I (purple line in Fig.~\ref{fig:PkModelI}). In the right panel we plot the numerical spectrum (orange line) and the fit with the analytical template in Eq.~\eqref{temp:delta}.    We also plot the sensitivity curves for LISA~\cite{Audley:2017drz,Caprini:2015zlo},  BBO and ultimate DECIGO~\cite{Yagi:2011wg}.  }
	\end{figure}

    \item {\em Resonant feature 1}. We are going to present two examples of the resonant feature of the standard clock signal. The purple spectrum shown in Fig.~\ref{fig:PkModelI} of Model I, which we also report in Fig.~\ref{fig:OmegaDelta}, is the first example.
    Here, the model is tuned such that most of the inflationary energy of the first stage of inflation is damped into the massive field oscillation when the massive field falls tachyonically. So we have a brief matter-dominated epoch between two stages of inflation.
    The features are produced when the massive field undergoes high frequency oscillations and the PPS consists of a series of very spiky peaks.  Like in the previous case, overall we see a broad peak followed by a narrow peak, although the narrow peak is sharper than the previous case because the resonance signal in the PPS is more spiky. However, on top of the narrow peak we see small oscillations whose frequency dependence is typical of the resonant running. This is the signature of the resonant feature generated by the oscillation of the massive field.
    
    Also, we note that the rising part of $\Omega_{\rm gw}$, before the bump, now shows a step instead of a simple power-law. Although this falls outside the sensitivity of LISA, it can be detected by BBO and DECIGO if the feature is shifted towards smaller scales, i.e. if the massive field starts to oscillate closer to the end of inflation.

    The observationally relevant part of the SGWB spectrum can be modeled analytically using the following template consisting in a broken power-law to model the broad structure of the bump and a log-normal peak with superimposed logarithmic oscillations to model the second peak together with features:

	\begin{equation}
\label{temp:delta}
h^2\Omega_{\rm GW}(x=f/f_p)= \begin{cases}
A_1\,\left(\frac{f}{f_1}\right)^{\alpha}&\text{for $f\leq f_1 $}\\
A_2\,\exp\left[-b_1(-\ln x -B_1)^\beta\right]&\text{for $f_1<f<f_2$}\\
A_3\Biggl\{\,\exp\left[-d_1\ln x\,-\sum_{i=2}^3d_i\left(\ln x - D_i\right)^{\delta_i}\right]+&\\
A_4\exp\left[- g_2\left(\ln x \right)^{\gamma_2}\right]g(f)\Biggr\}&\text{for $f>f_2$}.
\end{cases}
\end{equation}
where $f_p$ is the frequency of the narrow peak in Fig.~\ref{fig:OmegaDelta} and $g(f)=\left[1+l\sin \left(\omega\, \ln(f/f_p) + \phi\right)\right]$.
This frequency-dependent resonant running explicitly demonstrates the resonant feature present in the numerical results.
	
    	As shown in Fig.~\ref{fig:OmegaDelta}, the signal can be reproduced quite well by using the parameters given in Appendix~\ref{app:templates}.
    	
    	The full standard clock signal should also contain the sinusoidal feature generated by the sharp feature, as clearly seen in the PPS. However, in contrast to the last example, here the sinusoidal feature has a weaker amplitude, so it is the resonant feature that shows up in the SGWB spectrum.

			\begin{figure}
		\begin{center} 
			\resizebox{214pt}{172pt}{\includegraphics{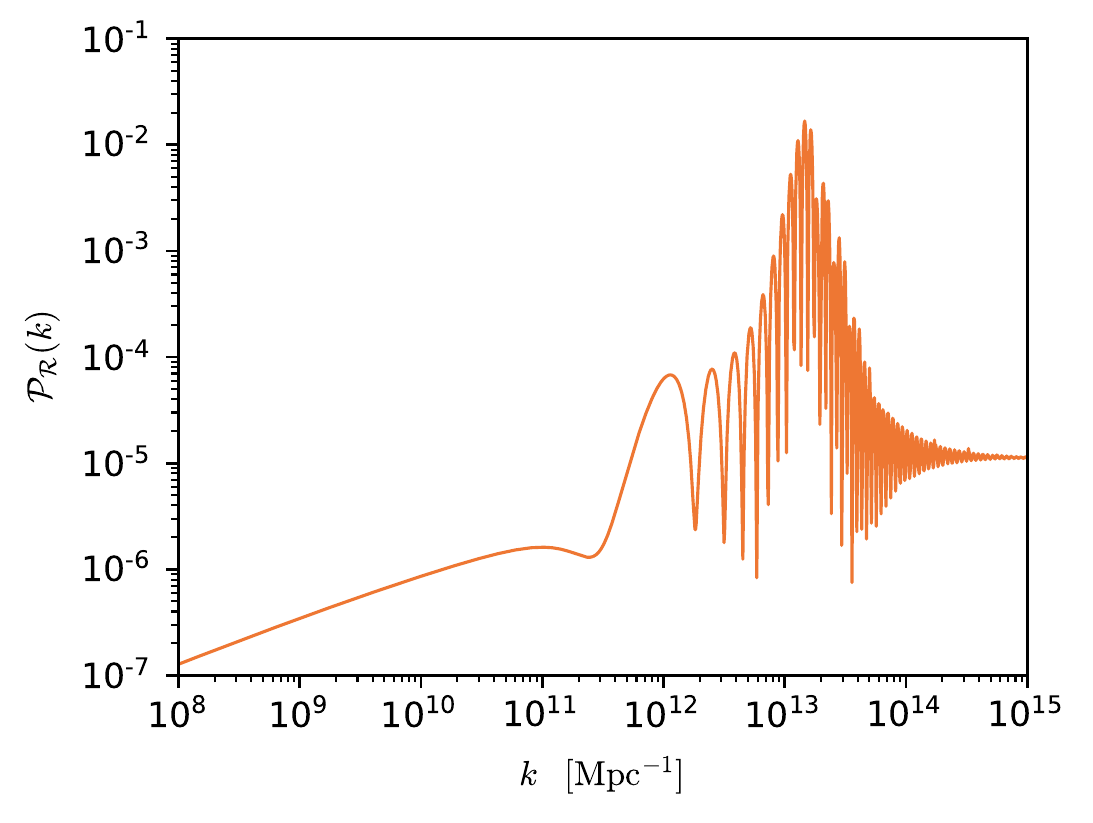}}
			\resizebox{214pt}{172pt}{\includegraphics{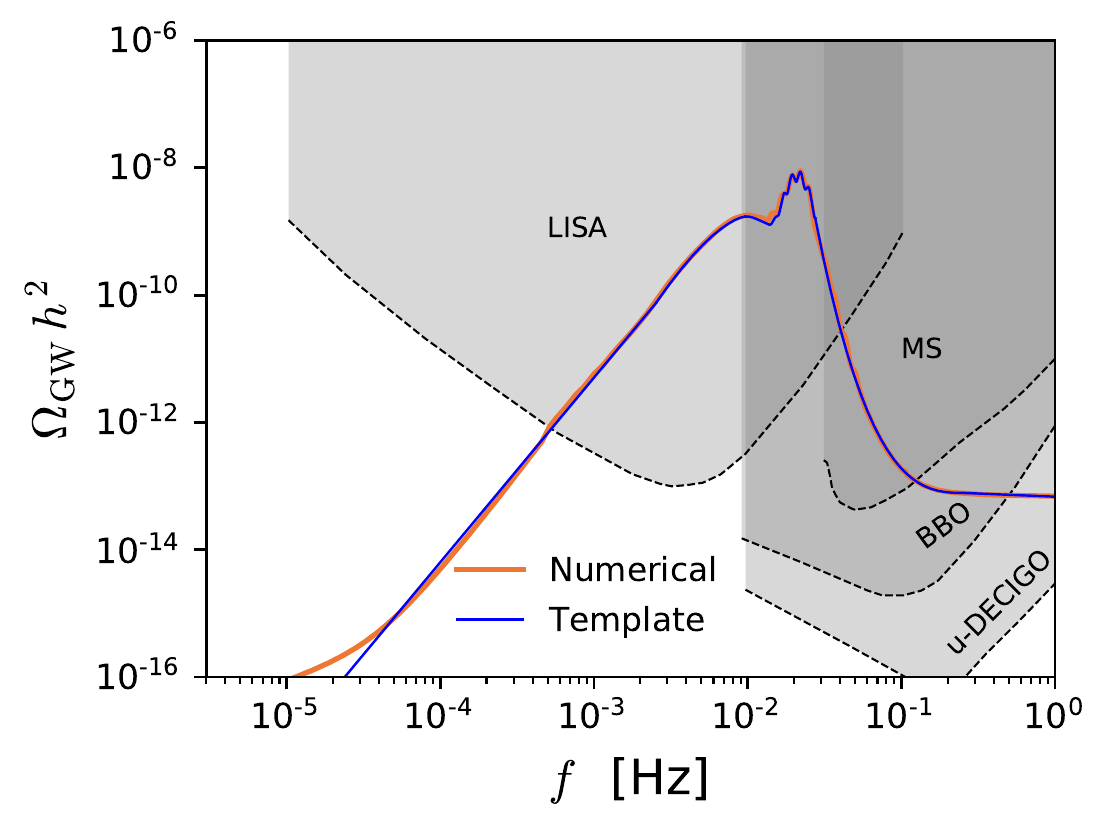}}	
		\end{center}
		\caption{\label{fig:OmegaMixed} Primordial scalar power spectrum [left] and SGWB energy density [right] for the resonant feature 2   produced in Model II (blue line in Fig.~\ref{fig:PkModelII}). In the right panel we plot the numerical spectrum (orange line) and the fit with the analytical template in Eq.~\eqref{temp:mixed}.    We also plot the sensitivity curves for LISA~\cite{Audley:2017drz,Caprini:2015zlo},  BBO and ultimate DECIGO~\cite{Yagi:2011wg}.  }
	\end{figure}

    \item {\em Resonant feature 2}. This is another example of resonant feature. In this example, during the sharp feature and massive field oscillation, the background evolution is still inflationary. This is different from the previous example, and closer to what happens in \cite{Chen:2014cwa,Chen:2014joa}. Comparing to the previous example, this example also has a different scale-dependence in its non-oscillatory part of the spectrum. 
    
    This example is produced only in Model II, and there is a suppression of the power spectrum at the scale-invariant CMB scales relative to the feature scales.
    Since we have to match large scales to the COBE normalization, this significantly raises the overall non-oscillatory power spectrum at the feature scales.  The light blue line shown in  Fig.~\ref{fig:PkModelII} of Model II, also reported in Fig.~\ref{fig:OmegaMixed}, is an example of this feature. 
    The appearance of the two peaks and the superimposed resonance-running ripples are very similar to the previous case, except that, around the time when this resonant feature is generated, the background is always inflationary.
    
    Away from the peaks, there are some very model-specific properties:  the rising part is smoother, since we have a smooth rise of the PPS from CMB scales to the ones at which it peaks. Furthermore, in the sensitivity range of BBO and DECIGO, $\Omega_{\rm GW}$ can be observed as a broken power-law. We emphasize that the tilt of the power-law encodes information on the field $\Theta$ driving the second stage of inflation.

    The structure of this signal is essentially the same as the previous case, and can be modeled analytically starting from the template in Eq.~\eqref{temp:delta}.  However, here we have a plateau with a large amplitude at small scales and, in order to model it, we need to add another domain to the template, which becomes:

	\begin{equation}
	\label{temp:mixed}
	h^2\Omega_{\rm GW}(x=f/f_p)= \begin{cases}
A_1\,\left(\frac{f}{f_1}\right)^{\alpha}&\text{for $f\leq f_1 $}\\
A_2\,\exp\left[-b_1(-\ln x -B_1)^\beta\right]&\text{for $f_1<f<f_2$}\\
A_3\Biggl\{\,\exp\left[-d_1\ln x\,-\sum_{i=2}^3d_i\left(\ln x - D_i\right)^{\delta_i}\right]+&\\
A_4\exp\left[- g_2\left(\ln x \right)^{\gamma_2}\right]g(f)\Biggr\}&\text{for $f_2<f<f_3$}\\
A_5\exp\left[\left(-\ln\frac{f}{f_4}\right)^\kappa\right]&\text{for $f_3<f<f_4$}\\
A_5\,\left(\frac{f}{f_4}\right)^{\iota}.
\end{cases}
\end{equation}
		As shown in  Fig.~\ref{fig:OmegaMixed}, the mixed signal can be reproduced quite well by using the parameters given in Appendix ~\ref{app:templates}.

\end{itemize}

The templates above can be summarized in a unified way as:

	\begin{equation}
\label{eq:template}
h^2\Omega_{\rm GW}(x=f/f_p)= \begin{cases}
A_1\,\left(\frac{f}{f_1}\right)^{\alpha}&\text{for $f\leq f_1 $}\\
A_2\,\exp\left[-b_1(-\ln x -B_1)^\beta\right]&\text{for $f_1<f<f_2$}\\
A_3\Biggl\{\,\exp\left[-d_1\ln x\,-\sum_{i=2}^3 d_i\left(\ln x - D_i\right)^{\delta_i}+d_4\, x\right]+&\\
A_4\exp\left[-g_1\ln x\,-\sum_{i=2}^3 g_i\left(\ln x - G_i\right)^{\gamma_i}\right]g(x)\Biggr\}&\text{for $f_2<f<f_3$}\\
A_5\exp\left[\left(-\ln\frac{f}{f_4}\right)^\kappa\right]&\text{for $f_3<f<f_4$}\\
A_5\,\left(\frac{f}{f_4}\right)^{\iota},
\end{cases}
\end{equation}
where 
\begin{equation*}
g(x)= \begin{cases}
0&\text{for bump features}\\
\left[1+l\sin \left(\omega\, (f-f_p) + \phi\right)\right] &\text{for sinusoidal features}\\
\left[1+l\sin \left(\omega \ln x + \phi\right)\right]&\text{for resonant features}.
\end{cases}
\end{equation*}
In the template above $f_p$ is the peak frequency in the bump feature, the frequency of the first peak in the sinusoidal ripple in the sinusoidal feature and the frequency of the narrow peak in the resonant features, respectively.
The template \eqref{eq:template} summarizes the main result of this paper.  Appendix \ref{app:templates} provides a visual illustration on how different domains of this template are patched together. If, for any reason, we are only interested in the main part of this GW spectrum, including only the top of the broad and narrow peaks as well as the oscillatory ripples, it is sufficient to only include parts of the template within the domain $f_1<f<f_2$ and $f_2<f<f_3$, and ignore the other parts.

\subsection{Comparison with other sources of SGWB}
\label{sec:comparison}
The previous Sections show that it is possible to produce a variety of spectral shapes of the SGWB within our model, though they can always be classified into three or four representative classes, for which we derived dedicated templates. Assuming that future GW experiments will be able to tell these classes apart, as the results of Section~\ref{sec:reconstruction} seem to suggest, a  natural question is whether it is possible to reproduce these classes within other models (see Ref.\cite{Caprini:2018mtu} for a review of sources of SGWB). 

The main property which is shared by all the four classes of features that we have introduced is the overall peaked shape of the $\Omega_{\rm GW}$. Some physical mechanisms that are known to  generate single or multi peaked structure of the SGWB are single-field inflationary models with a nearly inflection point~\cite{Bhaumik:2019tvl,Ballesteros:2020qam,Bhaumik:2020dor,Ragavendra:2020sop} or a non-canonical kinetic term~\cite{Lin:2020goi,Yi:2020cut,Zhang:2020uek}, certain variants of axion inflation with explosive production of gauge fields~\cite{Cook:2011hg,Barnaby:2011qe,Namba:2015gja,Garcia-Bellido:2016dkw,Ozsoy:2020ccy,Ozsoy:2020kat}, cosmic strings~\cite{Vachaspati:1984gt,Sakellariadou:1990ne,Hindmarsh:1990xi,Ringeval:2017eww,Matsui:2020hzi} and strong first order phase transitions during the thermal era of the Universe~\cite{Kosowsky:1992rz,Kamionkowski:1993fg,Damour:2000wa,Huber:2008hg}. Note that a peaked SGWB can also model the foregrounds given by unresolved sub-threshold mergers of galactic binaries~\cite{Cornish:2018dyw}.
 The spectral shape of the resulting SGWB, however, can be usually modeled with simple broken  power-laws~\cite{Caprini:2009fx,Caprini:2015zlo,Kuroyanagi:2018csn,Caprini:2019egz} or with a log-normal peak. This has to be contrasted with  the structure of the broad peak in our  templates,  which is rounded in the region of the peak, unlike a simple broken power-law where the bending is usually sharper, and, is not symmetric as it  presents a quite sharp fall after the peak going towards smaller scales, unlike a log-normal peak.

More complicated shapes can be produced by the process of preheating~\cite{GarciaBellido:2007dg,Dufaux:2007pt,GarciaBellido:2007af,Dufaux:2008dn,Dufaux:2010cf,Figueroa:2016ojl},  a small periodic structure upon the inflaton potential  \cite{Cai:2019amo,Cai:2019bmk}, oscillons at the end of inflation~\cite{Zhou:2013tsa,Antusch:2016con,Antusch:2017vga,Liu:2017hua,Amin:2018xfe} or phase transitions during inflation~\cite{An:2020fff,Ashoorioon:2020hln}. None of the models just mentioned, though, is able to reproduce the features observed in our four feature classes, especially those containing the oscillatory dependence that is characteristic of inflationary features.

The uniqueness of the signals produced in our models therefore allows us to probe features in the inflationary potential and oscillations of massive fields, and access scales completely different from those tested with CMB observations, opening a new window on the observation of primordial features from the primordial universe.

\section{Conclusions}
\label{sec:conclusions}

In this paper, we study the possibility of observing primordial feature signals from inflation models in the spaced-based GW experiments. In particular we make use of LISA sensitivities according to survey specifications. By varying choices of potentials and non-canonical couplings, we explore the phenomenology of a two-field inflation model, within the framework of a non-canonical two-field Lagrangian where the inflation is driven in two stages separately dominated by each of the fields.   According to the LISA observational window, we concentrate on the amplification of power spectra around $k\sim10^{10}-10^{13}~ \mathrm{Mpc}^{-1}$ and therefore the first phase of inflation is dominated by the first field till 40 {\it e-folds} before it starts to oscillate at the bottom of its potential, following what the second field takes over. 
Our model can be used to produce several main classes of features in different parts of its parameter space and we demonstrate that certain short-scale density perturbations need to be considerably boosted by fine-tuned primordial features in order for them to show up in the sensitivity regimes of the GW experiments.  Nonetheless, once they show up, some main characters of different types of features are kept and may be differentiated. One such character is that the features show a very steep growth that can significantly violate the single-field bound $n_s-1\leq4$. We numerically compute the profiles of various primordial features in terms of the spectrum of the SGWB. To study the prospects of observing these profiles in proposed observations, we discuss two approaches. The first approach is to reconstruct the signal from data, which requires very high signal-to-noise ratio. Using the binned Power-Law integrated Sensitivity formalism, we classify the signals according to the bin-width required for their high SNR reconstruction with LISA. Another more sensitive approach is to search for the analytical templates that represent the signals of primordial features. For the latter purpose, we are able to summarize all main properties of the GW profiles of several classes of primordial features in a template, given in Eq.~\eqref{eq:template}. This template schematically takes the following form:
\bea
\Omega_{\rm GW}(f) = {\rm broad~peak} + {\rm narrow~peak}~(1+ {\rm oscillatory~feature}) ~,
\eea
where the broad peak and narrow peak are generally present for SGWB generated by a spike in the scalar perturbation, but their overall shapes are modified due to the specific shape of spikes generated by primordial features. The more distinctive properties of primordial features are two types of oscillatory behaviors that show up on top of the narrow peak. In the case where only the first bump of the oscillations is significant, i.e.~the bump feature, only the broad peak is present.

There are many different ways the primordial universe may generated observable SGWB, as discussed in Sec.~\ref{sec:comparison}. It would be interesting to summarize these signals also in terms of analytical templates with characters that represent their distinguishable underlying physics. The final outcome of a template bank would be useful for future data analyses.

A final remark is in order. In our analysis, we assume a Gaussian statistics of primordial fluctuations. However, the  models introduced here,  like other field models with turns in the field space \cite{Garcia-Saenz:2018ifx,Garcia-Saenz:2019njm,Fumagalli:2019noh,Bjorkmo:2019aev}, can generate highly correlated higher order correlation functions and deviate from the pure Gaussian statistics \cite{Chen:2014joa,Chen:2014cwa}. It is known that  non-Gaussianities alter the process of PBH formation \cite{Young:2013oia,Garcia-Bellido:2017aan,Franciolini:2018vbk,Atal:2018neu,DeLuca:2019qsy,Yoo:2019pma,Ezquiaga:2019ftu}, which might invalidate the assumption of neglecting it in our paper. Even more importantly, non-Gaussianities can significantly change the spectral shape of the SGWB \cite{Cai:2018dig,Unal:2018yaa,Ragavendra:2020sop} possibly enhancing or washing out the oscillatory signal, which motivates further investigations in this direction.

%	\section*{Acknowledgements}

\medskip
\noindent	
\textbf{Note added:}
While this project was nearly complete, a related paper \cite{Fumagalli:2020nvq}, also studying the imprints of primordial features produced in multifield inflation on the SGWB, appeared on the arXiv. The two works have some overlap.

\appendix
\section{Equations for linear perturbations and numerical method}	
\label{app:numericalmethod}
In this Appendix, we give useful formulae governing the evolution of linear perturbations and describe the procedure that we use to numerically integrate them.
Here, we use a different gauge from Ref.~\cite{Chen:2014cwa}. Nevertheless, the two are equivalent.
	Since we do not have anisotropic stresses to linear order in our model the perturbed FRLW metric takes the following form in the Newtonian gauge \cite{Mukhanov:1990me}
	\begin{equation}
	\d s^2=-(1+2\Phi)\,\d t^2+a^2(t)\, (1-2\Phi)\,\d{\bm x}^2 \,,
	\end{equation}
	where $\Phi$ is the Bardeen potential characterizing the perturbations.
	
	In two-field inflationary models, it is convenient to decompose the scalar field perturbations, say  $\delta\phi$
	and $\delta\chi$, along directions that are parallel and orthogonal 
	directions to the trajectory in the field space~\cite{Gordon:2000hv}.\footnote{Note that to better connect in the literature, where $\sigma$ and $\theta$ denote the adiabatic field and the turning rate angle respectively, in this Appendix, we trade the names of the two scalar fields to $\phi$ and $\chi$. In order to use the equations in this Appendix to obtain the results presented in the main paper, one would have to make the substitution $\phi\to\sigma$ and $\chi\to \Theta$.}
	These correspond to the so called instantaneous adiabatic and isocurvature 
	perturbations and, defining for convenience $f(\phi)\equiv e^{b(\phi)}$, they are given by the expressions~\cite{DiMarco:2002eb}
	\begin{subequations}
		\label{eq:ai}
		\begin{eqnarray}
		\label{adi}
		\delta\sigma&=&\cos\theta\, \delta\phi
		+\sin\theta\, {\rm e}^b\, \delta\chi,\\
		\label{iso}
		\delta s&=&-\sin\theta\,\delta\phi
		+\cos\theta\, {\rm e}^b\, \delta\chi,
		\end{eqnarray}
	\end{subequations}
	with the quantity $\theta$ being the bending angle of the trajectory in the field space, 
	defined through the relations
	\begin{equation}
	\cos\theta=\frac{\dot{\phi}}{\dot{\sigma}},\quad
	\sin\theta={\rm e}^b \frac{\dot{\chi}}{\dot{\sigma}},\quad
	\dot{\sigma}^2=\dot{\phi}^2+{\rm e}^{2b}\dot{\chi}^2.
	\end{equation}
	Using the equations~(\ref{eq:sf}) one can  arrive at the following equations governing 
	$\dot{\sigma}$ and the angle $\theta$:
	\begin{subequations}
		\begin{eqnarray}
		\label{KGsigma}
		\ddot{\sigma}+3 H \dot{\sigma}+V_\sigma&=&0,\\
		\dot{\theta}&=&-\frac{V_s}{\dot{\sigma}}-b_\phi\dot{\sigma}\sin\theta
		\end{eqnarray}
	\end{subequations}
	where $V_\sigma$ and $V_s$ are given by 
	\begin{subequations}
		\begin{eqnarray}
		V_\sigma&=&V_\phi\cos\theta+{\rm e}^{-b}\,V_\chi\sin\theta,\\
		V_s&=&-V_\phi\sin\theta+{\rm e}^{-b}\,V_\chi\cos\theta.
		\end{eqnarray}
	\end{subequations}
	 We also introduce here the following quantities for future convenience
	\begin{subequations}
		\begin{eqnarray}
		V_{\sigma\sigma} &=&V_{\phi\phi}\cos^2\theta 
		+ {\rm e}^{-b} V_{\phi\chi} \sin 2 \theta
		+ {\rm e}^{-2 b} V_{\chi\chi} \sin^2\theta,\\
		V_{ss} &=&V_{\phi\phi}\sin^2\theta
		- {\rm e}^{-b} V_{\phi\chi}\sin 2 \theta
		+ {\rm e}^{-2 b} V_{\chi\chi}\cos^2\theta ,\\
		V_{\sigma s} &=&-V_{\phi\phi}\cos\theta\sin\theta
		+ {\rm e}^{-b} V_{\phi\chi} (\cos^2\theta-\sin^2\theta)
		+ {\rm e}^{-2 b} V_{\chi\chi}\cos\theta\sin\theta.
		\end{eqnarray}
	\end{subequations}

	To characterize the perturbations, it is customary to consider the gauge-invariant Mukhanov-Sasaki variable $Q_\sigma$ associated to $\delta \sigma$, i.e. $Q_\sigma = \delta \sigma + \dot \sigma/H \Phi$, and $\delta s$, already gauge invariant, associated with the adiabatic and isocurvature fields \cite{Lalak:2007vi}.
	The equations of motion governing $Q_\sigma$ and $\delta s$ are 
	\begin{subequations}
		\label{eq:Q}
		\begin{eqnarray}
		\ddot{Q}_\sigma + 3 H \dot{Q}_\sigma
		&+&\l[\f{k^2}{a^2}+V_{\sigma\sigma}-\dot{\theta}^2
		-\f{1}{a^3 \Mpl^2}
		\l(\f{a^3\dot{\sigma}^2}{H}\r)^{\fatdot}
		+b_\phi u(t)\r]Q_\sigma\nn\\
		&=& 2\l(\dot{\theta} \delta s \r)^{\fatdot}
		-2\left(\frac{\dot{H}}{H}
		+\frac{V_\sigma}{\dot{\sigma}}\right)\dot{\theta} \delta s
		+  b_{\phi\phi}\dot{\sigma}^2\sin2\theta \delta s +2 b_\phi h(t)  \,,\qquad\\
		\ddot{\delta s} + 3 H\dot{\delta s}
		&+&\l[\f{k^2}{a^2}+m_\textup{iso}^2\r] \delta s
		= 2\f{V_s}{H}\l(\f{H}{\dot{\sigma}}Q_\sigma\r)^{\fatdot},
		\end{eqnarray}
	\end{subequations}
	where
 \begin{equation}
 	\label{eq:effmassiso}
m_\textup{iso}^2\equiv V_{ss}+3\dot{\theta}^2
+b^2_\phi g(t)+b_\phi f(t)
-b_{\phi\phi}\dot{\sigma}^2
-4\f{V_s^2}{\dot{\sigma}^2},
\end{equation}
	and	the quantities $u(t)$, $h(t)$, $g(t)$ and $f(t)$ are given by \cite{DiMarco:2002eb}
	\begin{subequations}
		\begin{eqnarray}
		u(t)&=&\dot{\theta}\dot{\sigma}\sin\theta-{\rm e}^{-b}V_\chi\sin\theta\cos\theta,\\
		h(t)
		&=&-\dot{\sigma} (\sin\theta\, \delta s)^{\fatdot}
		-\sin\theta \left(\frac{\dot{H}}{H}\dot{\sigma}
		+2 V_\sigma\right) \delta s-3 H\dot{\sigma}\sin\theta \delta s,\\
		g(t)&=&-\dot{\sigma}^2(1+3\sin^2\theta),\\
		f(t)&=&V_\sigma(1+\sin^2\theta)-4 V_s\sin\theta.
		\end{eqnarray}
	\end{subequations}

	In order to numerically integrate the equations above, we impose the following Bunch-Davies 
	initial conditions on the variables $Q_\sigma$ and $\delta s$ 
	at very early times when the modes of cosmological interest are well 
	inside the Hubble radius:
	\begin{equation}
	\label{eq:ic}
	Q_\sigma(\tau)
	\simeq \delta s(\tau)
	\simeq \f{1}{a(\tau)}\,\f{{\rm e}^{\imath k \tau}}{\sqrt{2 k}},
	\end{equation}
	where $\tau$ is the conformal time coordinate defined as $\tau=
	\int\, \d t/a(t)$. It was verified in Ref.~\cite{Braglia:2020fms}  that the same results are obtained by considering Bunch-Davies 
	initial conditions for the gauge-invariant variables associated to $\phi$ and $\chi$ instead. Extra care is however needed when imposing the initial conditions in case the turning rate $\dot{\theta}$ is very large at early times so that $Q_\sigma$ and $\delta_s$ can be initially correlated \cite{Cremonini:2010ua}.
	
	While $Q_\sigma$ and $\delta s$ are convenient variables to identify the most natural initial conditions that can be imposed, it is also useful to think in terms of curvature and 
	isocurvature fluctuations, which are used to classify initial conditions in the post-inflationary expansion.
		These quantities are given in terms of the variables $Q_\sigma$ and $\delta s$ by the following relations:
	\begin{equation}
	\mathcal{R}= \frac{H}{\dot{\sigma}}Q_\sigma,\quad
	\mathcal{S}=\frac{H}{\dot{\sigma}} \delta s.
	\end{equation}

	It is straightforward
	to arrive at the following equations governing the curvature
	and the isocurvature perturbations $\cR$ and $\cS$:
	\begin{subequations}
		\label{eq:RS}
		\begin{eqnarray}
		\label{eq:R}
		\ddot{\mathcal{R}}
		&+&\left(H+2\f{\dot{z}}{z}\r)\dot{\mathcal{R}}
		+\frac{k^2 }{a^2}\mathcal{R}
		=-\frac{2 V_s}{\dot{\sigma}}\dot{\mathcal{S}}
		-\,2\biggl(-{\rm e}^{-b}  b_{\phi }\cos^2\theta\, V_{\chi }
		+ \sin\theta b_{\phi } V_{\sigma}\nn\\
		& &\qquad\qquad\qquad\qquad\qquad\;\;
		+\, V_{\sigma s}+\f{\dot{\sigma}}{H M^2_\textup{pl}}V_s\biggr)\mathcal{S},\\
		\ddot{\mathcal{S}}
		&+&\l(H+2\f{\dot{z}}{z}\right)\dot{\mathcal{S}}
		+\biggl\{\frac{k^2}{a^2}-2 H^2-\dot{H}+\frac{H \dot{z}}{z}
		+\frac{\ddot{z}}{z}-\dot{\theta}^2-\dot{\sigma}^2 b_{\phi }^2 \cos^2\theta
		-\dot{\sigma}^2 b_{\varphi \varphi }+V_{ss}\nn\\
		&+&\,b_{\phi } \l[4 \sin \theta \, V_s
		+(1+\sin^2\theta) V_{\phi }\r]
		\biggr\}\cS
		=\f{2  V_s}{\dot{\sigma}}\dot{\mathcal{R}},
		\end{eqnarray}
	\end{subequations}
	where $z\equiv a \dot{\sigma}/H$.

Our numerical integration proceeds as follows. First, we  
first integrate the equations~(\ref{eq:sf}) and~(\ref{eq:f}) to 
 get the background quantities, working with the numer of $e$-folds as the time variable. 
In all the models considered, we  assume 
that the pivot scale $k=0.05 \,\text{Mpc}^{-1}$ leaves the Hubble 
radius at $50$ e-folds before the end of inflation. 

With the background quantities in hand, we  then integrate 
the equations~(\ref{eq:RS}) governing the curvature and isocurvature perturbations~$\cR$ and~$\cS$.
Following Refs.~\cite{Tsujikawa:2002qx,Lalak:2007vi}, we  integrate the equations~(\ref{eq:RS}) by imposing the Bunch-Davies initial condition~(\ref{eq:ic}) on $Q_\sigma$ and 
assuming the initial value of $\delta s$ to be zero.
We  then integrate the equations a second time by interchanging the initial conditions on~$Q_\sigma$ and~$\delta s$. This ensures no correlations between the curvature and isocurvature fluctuations when the modes are deep inside the Hubble radius. 
If we denote these two sets of solutions as ($\cR_{1}$, $\cS_1$) 
and ($\cR_2$, $\cS_2$), then the power spectra describing the curvature and the isocurvature perturbations as well as the cross-correlations between them are defined respectively as
\begin{subequations}
	\label{P1P2}
	\begin{eqnarray}
	\mathcal{P}_{\cR}(k)
	&=&\frac{k^3}{2\pi^2}
	\l(\lvert\cR_1\rvert^2+\lvert\cR_2\rvert^2\r)
	=\mathcal{P}_{\cR_1}(k)+\mathcal{P}_{\cR_2}(k),
	\label{eq:PR}\\
	\mathcal{P}_{\cS}(k)
	&=&\frac{k^3}{2\pi^2}
	\l(\lvert\cS_1\rvert^2+\lvert\cS_2\rvert^2\r),\\
	\mathcal{C}_{\cR\cS}(k)
	&=&\frac{k^3}{2\pi^2}
	\l(\cR^\ast_1\cS_1+\cR^\ast_2\cS_2\r).
	\end{eqnarray}
\end{subequations}
In terms of the in-in formalism, this procedure effectively sums up all tree-level diagrams non-perturbatively \cite{Chen:2015dga}.

\section{Matching the domains in the templates}
\label{app:templates}

In this Appendix, we report the parameters used in Figs.~\ref{fig:OmegaBump}, \ref{fig:OmegaWiggly}, \ref{fig:OmegaDelta} and \ref{fig:OmegaMixed} to match the template in Eq.~\ref{eq:template} to the correspondent $\Omega_{\rm GW}$ computed numerically.

\begin{itemize}
    \item {\em Bump feature.} The parameters that we use to reproduce Fig.~\ref{fig:OmegaBump} are:
 $A_1=8.82\times 10^{-10}$, $A_3=1.311\times10^{-7}$, $f_1=6.67\times 10^{-5} $ Hz, $f_p=1.73\times10^{-3}$ Hz, $\alpha=2.95$ and $d_1=d_4=-d_2=-0.36$.
\item {\em Sinusoidal feature.} The parameters that we use to reproduce Fig.~\ref{fig:OmegaWiggly} are:
$A_1=2.05\times10^{-10},\,$ $f_1=2.68\times10^{-4}$ Hz, $\alpha=3.05,\,$ $A_2=1.07\time10^{-8},\,$ $f_p=2.69\times 10^{-3}$ Hz,\, $f_2=1.2\times10^{-3}$ Hz, $b_1=2.02,\,$ $B_1=0.78,\,$ $\beta=1.6,\,$
$A_3=3.76\times 10^{-9},\,$ $A_4=11.02,\,$ $d_1=1.815,\,$ $d_2=0.8,\,$ $D_2=0.1,\,$ $\delta_2=6,\,$ $g_1=1.60,\,$ $g_2=18,\,$ $G_2=0.37,\,$ $\gamma_2=6,\,$ $g_3=0.8,\,$ $G_3=0.1,\,$ $\gamma_3=8,\,$ $l=0.3,\,$ $\omega=3600$.
\item {\em resonant feature 1.} The parameters that we use to reproduce Fig.~\ref{fig:OmegaDelta} are: 
$A_1=4.91\times10^{-11},\,$ $f_1=8.78\times10^{-4}$ Hz, $\alpha=3,\,$
 $A_2=1.07\time10^{-8},\,$ $f_p=7.43\times 10^{-3}$ Hz,\,  $f_2=3.33\times10^{-3}$ Hz, $b_1=2.02,\,$ $B_1=0.80,\,$ $\beta=1.6,\,$
$A_3=3.81\times 10^{-10},\,$ $A_4=12.87,\,$ $d_1=1.75,\,$ $d_2=0.1,\,$ $D_2=-0.56,\,$ $\delta_2=2\,$,$d_3=3.7,\,$ $D_3=-0.29,\,$ $\delta_3=4\,$  $g_2=23.26,\,$  $\gamma_2=6,\,$ $l=0.22,$ $\omega=48,\,$ $\phi=-1.17$.

\item {\em resonant feature 2.} The parameters that we use to reproduce Fig.~\ref{fig:OmegaMixed} are:$A_1 = 7.6\times10^{-11},\,$
$f_1=2.55\times 10^{-3}\,$ Hz,
$\alpha=2.9,\,$
$A_2= 1.71\times10^{-9},\,$
$g_2 = 22.2,\,$
$\gamma_2 = 2,\,$
$f_p = 0.021$ Hz,
$b_1 = 2.02,\,$
$B_1 = 0.78,\,$
$\beta = 1.5,\,$
$f_2 = 9.66\times10^{-3},\,$
$A_3 = 5.6\times10^{-10},\,$
$A_4 = 12.48,\,$
$d_2 = 0.10,\,$
$D_2 = 0.54,\,$
$\delta_2 = 2,\,$
$d_4 = 3.7,\,$
$D_4 = 0.27,\,$
$\delta_4= 4,\,$
$l = 0.22,\,$
$\omega = 48,\,$
$\phi= -2.01,\,$
$A_5 = 7.8\times10^{-14},\,$
$\kappa= 2.835,\,$
$\iota = -0.1,\,$
$f_4=0.0275,\,$
$f_5 = 0.261$.

\end{itemize}

\begin{figure}
		\begin{center} 
			\resizebox{214pt}{172pt}{\includegraphics{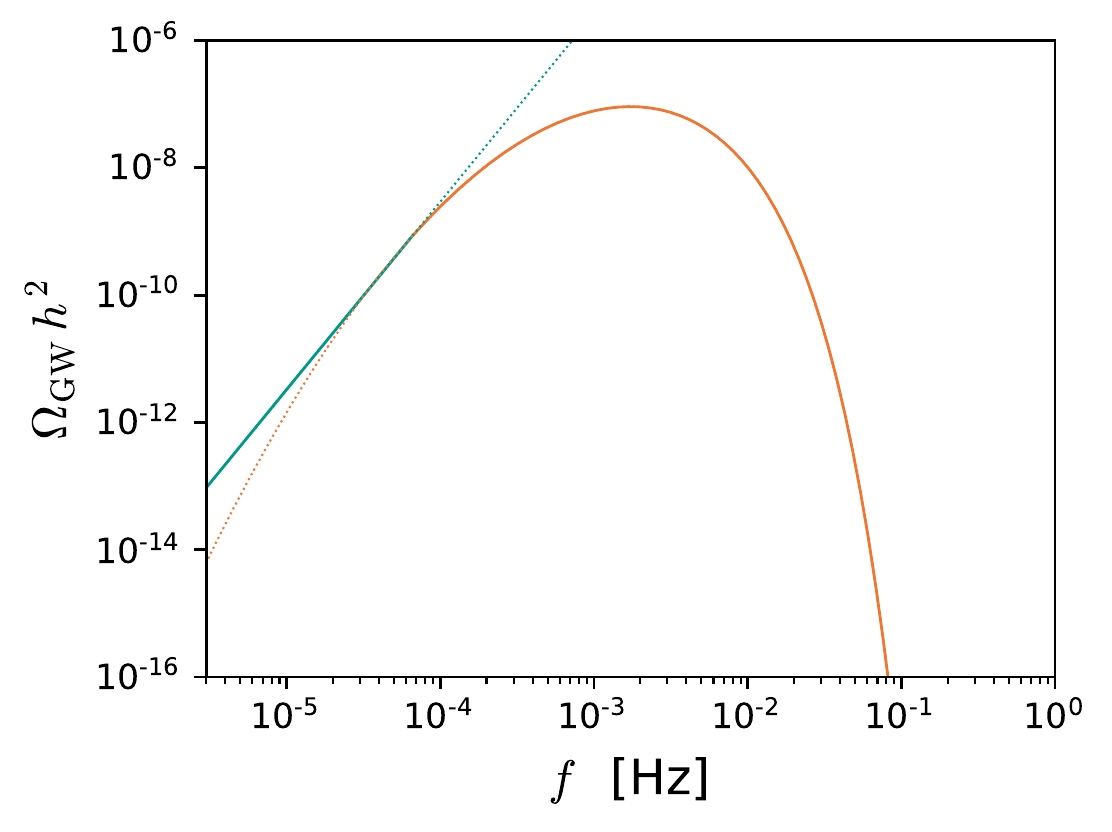}}
			\resizebox{214pt}{172pt}{\includegraphics{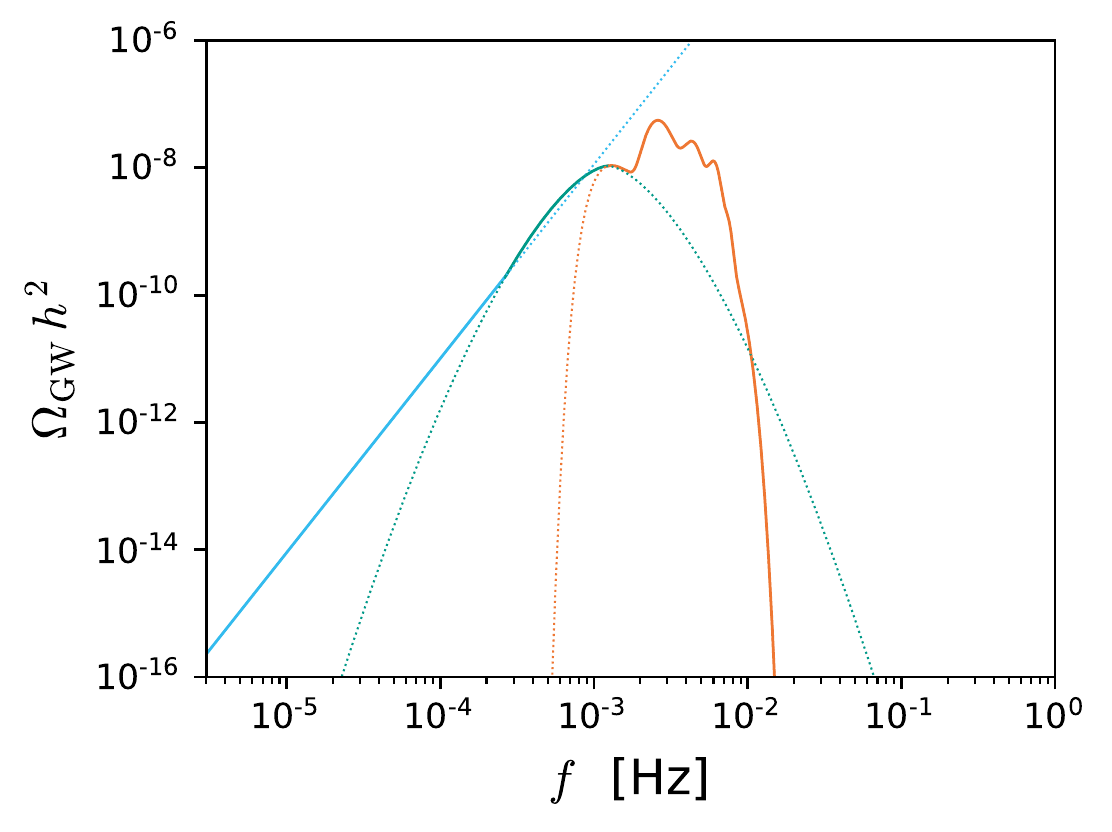}}
			\resizebox{214pt}{172pt}{\includegraphics{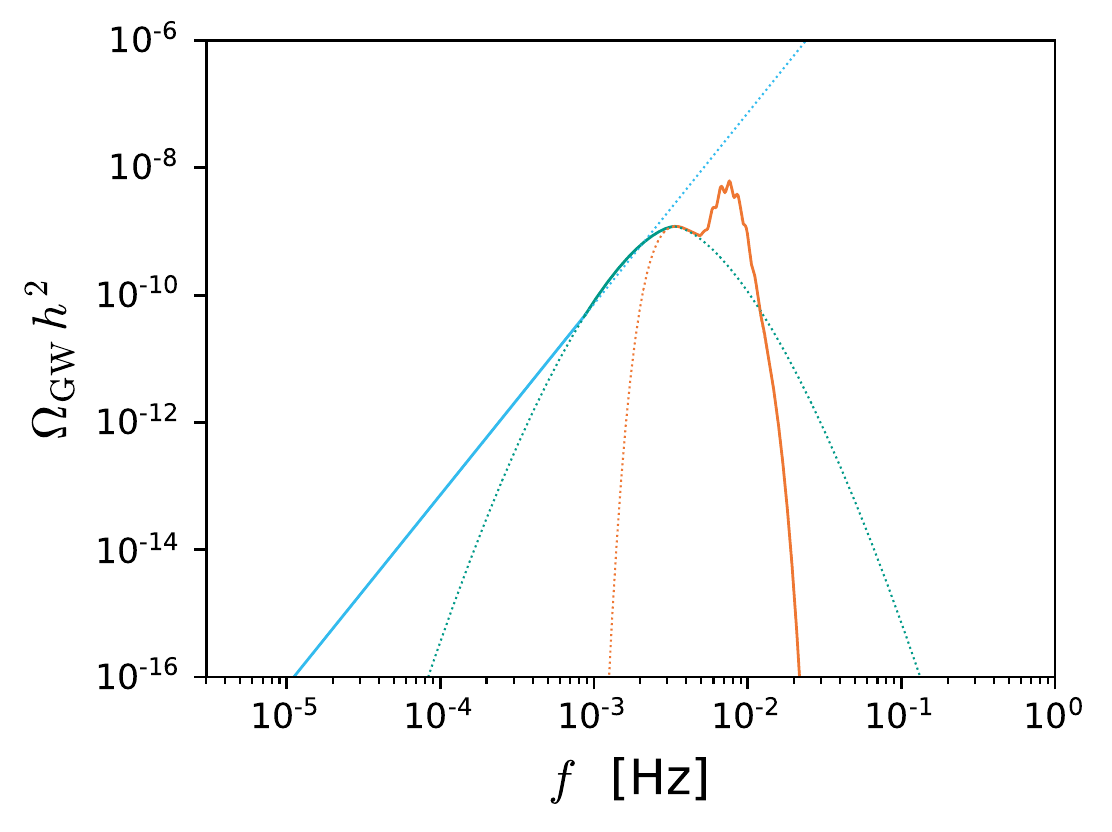}}
			\resizebox{214pt}{172pt}{\includegraphics{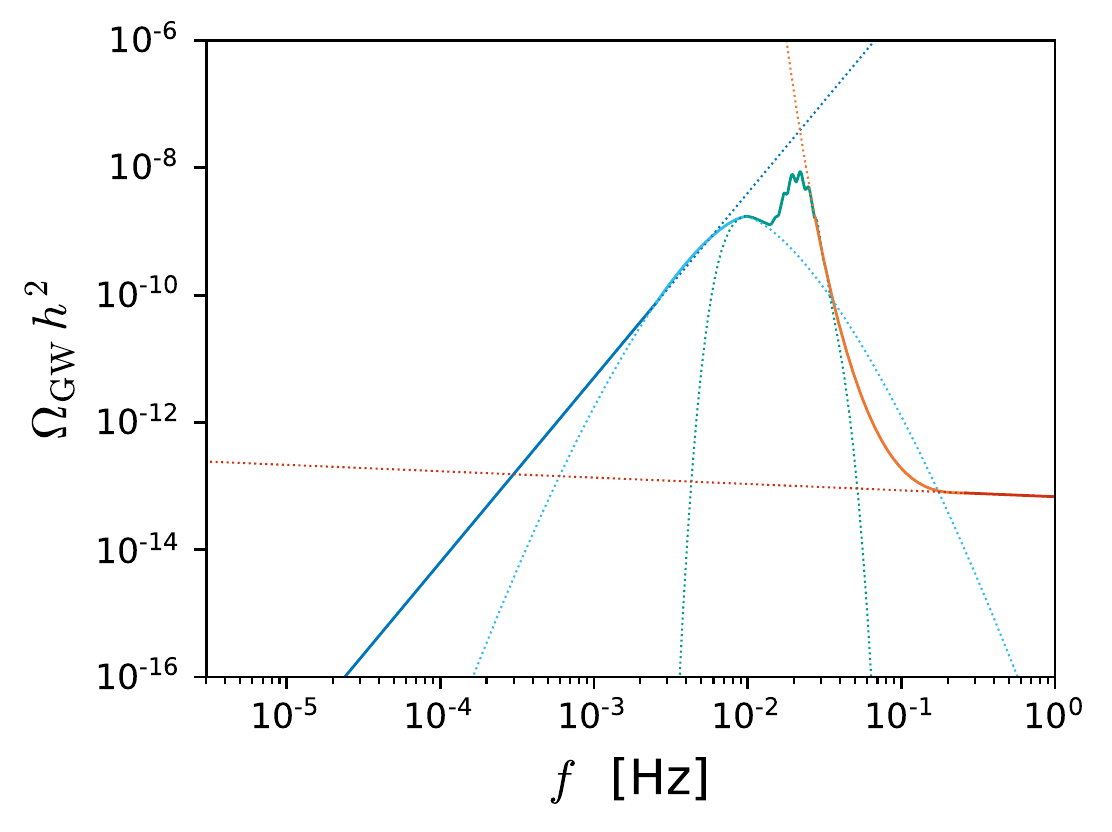}}	
		\end{center}
		\caption{\label{fig:templatesmatchinf} 
		We plot our templates explicitly showing the contribution of each of the domains in which they are divided. }
	\end{figure}

Furthermore, in Fig.~\ref{fig:templatesmatchinf} we demonstrate how the domains in Eq.~\eqref{eq:template} are divided, in order to guide the reader how our template is patched together. 
	
%	\section*{Acknowledgements}w

%	\appendix
%	\section{}\label{}	

	%%%%%%%%%%%%%%%%%%%%%%%%%%%%%%%%%%%%%%%%%%%%%%%%%%%%%%%%%%%%%%%%%%%%%%%%%%%%%%%
	\bibliographystyle{JHEP}
	\bibliography{SGWB}
	%%%%%%%%%%%%%%%%%%%%%%%%%%%%%%%%%%%%%%%%%%%%%%%%%%%%%%%%%%%%%%%%%%%%%%%%%%%%%%%
\end{document}